\documentclass[twocolumn]{aastex62}

\received{September 2019}
\revised{December 2019}
\accepted{January 2019}
\submitjournal{ApJ}

\usepackage{longtable}
\usepackage{amssymb}
\usepackage{color}
\usepackage{wrapfig}
\usepackage{amsmath}
\usepackage{longtable}
\usepackage[version=3]{mhchem}
\usepackage[normalem]{ulem}
\usepackage{booktabs}
\usepackage{graphicx}

\newcommand*{\Scale}[2][4]{\scalebox{#1}{$#2$}}%
\newcommand{\feh}{$\mbox{[Fe/H]}$}
\newcommand{\Lagr}{\mathcal{L}}

\begin{document}
\shorttitle{Photometric metallicities of Tucana II members} 
\title{Stellar metallicities from SkyMapper photometry I: A study of the Tucana II ultra-faint dwarf galaxy}

\correspondingauthor{Anirudh Chiti}
\email{achiti@mit.edu}

\author{Anirudh Chiti}
\affil{Department of Physics and Kavli Institute for Astrophysics and Space Research, Massachusetts Institute of Technology, Cambridge, MA 02139, USA \\}

\author{Anna Frebel}
\affil{Department of Physics and Kavli Institute for Astrophysics and Space Research, Massachusetts Institute of Technology, Cambridge, MA 02139, USA \\}

\author{Helmut Jerjen}
\affil{Research School of Astronomy and Astrophysics, Australian National University, Canberra, ACT 2611, Australia}

\author{Dongwon Kim}
\affil{Astronomy Department, University of California, Berkeley, CA, USA}

\author{John E. Norris}
\affil{Research School of Astronomy and Astrophysics, Australian National University, Canberra, ACT 2611, Australia}

\begin{abstract}

We present a study of the ultra-faint Milky Way dwarf satellite galaxy Tucana II using deep photometry from the 1.3\,m SkyMapper telescope at Siding Spring Observatory, Australia.
The SkyMapper filter-set contains a metallicity-sensitive intermediate-band $v$ filter covering the prominent Ca II K feature at 3933.7\,\AA.
When combined with photometry from the SkyMapper $u, g$, and $i$ filters, we demonstrate that $v$ band photometry can be used to obtain stellar metallicities with a precision of $\sim0.20$\,dex when [Fe/H] $> -2.5$, and $\sim0.34$\,dex when [Fe/H] $< -2.5$.
Since the $u$ and $v$ filters bracket the Balmer Jump at 3646\,\AA, we also find that the filter-set can be used to derive surface gravities.
We thus derive photometric metallicities and surface gravities for all stars down to a magnitude of $g\sim20$ within $\sim$75 arcminutes of Tucana II.
Photometric metallicity and surface gravity cuts remove nearly all foreground contamination.
By incorporating \textit{Gaia} proper motions, we derive quantitative membership probabilities which recover all known members on the red giant branch of Tucana II.
Additionally, we identify multiple likely new members in the center of the system and candidate members several half-light radii from the center of the system.
Finally, we present a metallicity distribution function derived from the photometric metallicities of likely Tucana II members.
This result demonstrates the utility of wide-field imaging with the SkyMapper filter-set in studying UFDs, and in general, low surface brightness populations of metal-poor stars.
Upcoming work will clarify the membership status of several distant stars identified as candidate members of Tucana II.

\end{abstract}

\section{Introduction}
\label{sec:introduction}

The Milky Way's satellite dwarf galaxies test paradigms of the formation and evolution of the local universe.
These systems are thought to be similar to those that were accreted to form the old Milky Way halo \citep{fn+15}.
Consequently, studying the stellar content of these ancient dwarf galaxies and comparing their stars to those in the Milky Way halo can probe potential connections between the these stellar populations \citep[e.g.,][]{ksg+08, fks+10}.
The relatively simple nature of dwarf galaxies also enables the modeling of their early chemical evolution \citep[e.g.,][]{kls+11, vsi+12, rbs+15, ewk+18} and the nature and properties of the earliest nucleosynthesis events \citep[e.g.,][]{jfc+16}.

Of particular interest in this regard are the Milky Way's ultra-faint dwarf galaxies (UFDs).
These systems have stellar populations that are old ($\gtrsim$ 10 Gyr) and metal-poor (average [Fe/H] $\lesssim -2.0$) (see \citealt{s+19} for a review), where a metal-poor star is defined as having an iron abundance $\feh\,\,\le -1$, in which [Fe/H] = $\log_{10}(N_{\text{Fe}}/N_{\text{H}})_{\star} - \log_{10}(N_{\text{Fe}}/N_\text{H})_\sun$. 
Hence, UFDs are particularly interesting targets both from the perspective of chemical evolution, since they are thought to have undergone only a few cycles of chemical enrichment and star formation, and from a cosmological perspective, since at least some of them are hypothesized to be surviving first galaxies \citep{fb+12}.

However, the faintness of UFDs makes it difficult to study their stellar population in detail.
Each system has only a handful of stars that are bright enough ($V \lesssim 19$) to obtain detailed chemical abundances.
Thus, the number of foreground stars generally outnumbers the number of bright(er) UFD stars in images of the galaxy.
This makes identifying UFD member stars for follow-up observations time-consuming, since stars along the giant branch in a color-magnitude diagram (CMD) must first be spectroscopically followed up with low or medium-resolution spectroscopy to measure velocities (and metallicities) to identify true member stars. 
Only for those confirmed member stars it is useful to obtain reliable chemical abundances, usually from high-resolution spectroscopy.

One can principally bypass the time-intensive intermediate step of low or medium-resolution spectroscopy by deriving metallicities from photometry, since UFD stars have collectively been shown to be metal-poor ([Fe/H] $\lesssim -2.0$), and thus generally more metal-poor than foreground halo stars \citep{abj+13}.
Indeed, \citet{pl+19} demonstrated that one can increase the efficiency in identifying member stars of UFDs by using metallicity-sensitive colors in Dark Energy Survey photometry.
Deriving reliable metallicities of individual metal-poor stars from photometry is a relatively recent techique \citep[i.e.,][]{smy+17}, building on previous studies that demonstrated that photometry could be used to identify metal-poor stars \citep{tlp+91}.
Photometric metallicities are generally computed by using a narrow-band imaging filter that is sensitive to the overall metallicity of the star due to the presence of a prominent metal absorption feature  (i.e., the Ca II K line) within the bandpass of the filter \citep[e.g.,][]{ksb+07, smy+17, wpb+19}.
Photometry has the additional benefit of being able to provide information on all stars within the field of view of the camera, whereas in spectroscopy, one is limited by e.g., slit arrangements, number of fibers, or pixels in the CCD mosaic.

The SkyMapper Southern Sky Survey \citep{ksb+07,wol+18} pioneered the search for metal-poor stars using a filter set that contains an intermediate-band $\sim300$\,{\AA} wide $v$ filter that encompasses the Ca II K line within its bandpass, making the flux through the $v$ filter dependent on the overall stellar metallicity. This narrow-band $v$ filter has already been used to identify a number of extremely metal-poor stars and several stars with [Fe/H] $< -6$ \citep{kbf+14, jkf+15, nbd+19}. Recent work from the Pristine Survey \citep{smy+17}, which uses a narrow-band filter centered on the Ca II K line at 3933.7\AA\,\, to obtain photometric metallicities, has been successful in applying this technique to find halo metal-poor stars \citep[e.g.,][]{ysa+17, sab+18} and also study UFDs to derive their metallicity distribution functions \citep{lms+18, lms+19}.

A goal of this paper is to demonstrate that the SkyMapper filter set \citep{bbs+11} can be used to chemically characterize UFDs via photometric measurements of stellar parameters. Besides metallicities from the $v$ filter, the relative flux through the SkyMapper $u$ and $v$ filters is sensitive to the surface gravity ($\log g$) of stars since those filters surround the Balmer Jump at 3646\,\AA\,\,\citep[e.g.,][]{mks+11}.
The $\log g$ of stars is of additional use as a discriminant when studying UFDs, since their member stars that are bright enough for spectroscopy are generally on the red giant branch (RGB) and should thus have low surface gravities.
We note that we do not analyze horizontal branch stars in this study, as it is difficult to discriminate their metallicities due to their relatively high effective temperatures.
Given this  difficulty in discriminating metallicities, photometry with higher precision than the data presented in this paper would be required to derive photometric metallicities for stars on the horizontal branch of Tucana II (g$\sim$19.2).

The Tucana II UFD was discovered in the Dark Energy Survey \citep{kbt+15, bdb+15}. 
Tucana II is relatively nearby (57\,kpc) and has a half-light radius of 9.83$\arcmin$ \citep{kbt+15}.
It was confirmed as a UFD by \citet{wmo+16}, who measured a large velocity dispersion ($\sigma_{\text{v}_{\text{los}}} = 8.6$\,km\,s$^{-1}$), a low mean metallicity of $\langle$[Fe/H]$\rangle$ $= -$2.23, and identified eight stars as probable members. 
\citet{jfe+16} and later \citet{cfj+18} presented chemical abundances from high-resolution spectroscopy of seven stars in Tucana II.
Interestingly, two of the stars from the \citet{cfj+18} sample were new member stars that were approximately two half-light radii from the center of Tucana II.
These stars were originally selected for spectroscopic follow-up based on the SkyMapper photometry described here. If not for these data, they would likely have been missed by traditional low or medium resolution spectroscopic selection techniques.

We obtained deep images (down to $g\sim22$) of the Tucana II UFD using the  $u,v,g$, and $i$ filters on the 1.3\,m SkyMapper telescope to demonstrate that we can 1) use the photometry to efficiently identify bright members for follow-up high resolution spectroscopy, and 2) use the photometric metallicities of the member stars to derive a metallicity distribution function (MDF) of dwarf galaxies. Since UFD members should have similar proper motions, we also use \textit{Gaia} DR2 proper motion data \citep{gaia+16, gaia+18} to further improve our selection of likely members.
As shown in e.g., \citet{pl+19} and \citet{cf+19}, \textit{Gaia} proper motion data is especially useful in removing foreground contaminants when studying dwarf galaxy member stars.

The paper is arranged as follows. In Section~\ref{sec:observations}, we discuss the observations, data reduction procedure, and precision and depth of our photometry; in Section~\ref{sec:grid}, we discuss generating a grid of synthetic photometry which we later use to derive stellar parameters; in Section~\ref{sec:analysis}, we outline our derivation of photometric metallicities and photometric $\log g$ values, and discuss our sources of uncertainty; in Section~\ref{sec:TucII}, we speculate on properties of Tucana II (i.e., the MDF) from our analysis; in Section~\ref{sec:conclusion}, we summarize our results.

\section{Observations \& Data Reduction}
\label{sec:observations}

\subsection{Photometry}
\label{sec:phot}

Observations of the Tucana II UFD were taken between July 19, 2015 and December 15, 2015 with the 1.35\,m SkyMapper telescope at Siding Springs Observatory, Australia, as part of an auxillary program to obtain deep photometry of UFDs.
Table~\ref{tab:photobs} summarizes our photometric observations. The SkyMapper camera has 32 4k$\times$2k CCD chips covering a 2.34$^{\circ}$ by 2.40$^{\circ}$ field of view, which enabled the imaging of the entire Tucana II UFD \citep[$r_{1/2} \sim$ 10\arcmin;][]{kbt+15, bdb+15} in each frame.
Images were taken with the custom SkyMapper $u, v, g,$ and $i$ filters \citep{bbs+11}.

We developed a data reduction pipeline for these data separately from the one used by the SkyMapper collaboration \citep{wol+18}, since no pipeline existed when the data for this project were collected in late 2015. 
The reduction procedure for SkyMapper data is not trivial due to the presence of systematic pattern-noise signatures in the raw data that, if not properly removed, would impair any photometric measurements, as can be seen in Figure~\ref{fig:patternnoise}.
We therefore explicitly outline each step of our data reduction procedure in Section~\ref{sec:photdata} and describe our handling of the pattern-noise removal process in Section~\ref{sec:photpattern}.
We then discuss the precision and completeness of our photometry in Section~\ref{sec:photcomplete}.

\subsubsection{Data Reduction}
\label{sec:photdata}

Data reduction was mostly performed following standard procedures \citep{h+06} using python scripts that utilized the \texttt{astropy} package \citep{astropy}.
Bias-subtraction was done on a row-by-row basis using the overscan region that was 50 pixels wide.
Flat-field corrections were applied using master flat-field frames that were generated for each filter on each night data was taken.
Each master flat-field frame was generated by median-filtering 5 to 10 individual frames.
We note that we used the master flat-field frames from July 20, 2015 for the $u$ and $v$ data on July 26, 2015, since an insufficient number of individual flat-field frames were obtained for those filters on that date.

Then, we derived astrometric solutions for each frame following a two-step procedure. 
First, an estimate of the astrometric solutions was computed using \texttt{astrometry-net} \citep{lhm+10}. 
These astrometric solutions were subsequently corrected with SCAMP \citep{b+06}. 

We derived photometric zero-point corrections for each of our exposures by comparing our measured instrumental magnitudes to calibrated magnitudes in the public SkyMapper DR1.1 catalog (released on December 2017) that were derived with the analysis pipeline used by the SkyMapper collaboration \citep{wol+18}.
We first compiled source catalogs for each of our exposures using the default configuration of the Source Extractor program \citep{ba+96} except for the following parameters to ensure appropriate background subtraction: \texttt{BACKPHOTO\_TYPE=LOCAL}, \texttt{BACKSIZE=100,100}, and \texttt{BACKPHOTO\_THICK=10}.
Then, for each exposure, we cross-matched our source catalog to the public SkyMapper DR1.1 catalog and derived a zeropoint offset for our magnitudes by taking the weighted average of the difference between our measured instrumental magnitudes and those in the SkyMapper DR1.1 catalog.
To improve the precision of our measured offsets, we only compared the magnitudes of unblended stars brighter than 16th magnitude, and compared our aperture photometry, based on the MAG\_AUTO keyword in Source Extractor, to the aperture photometry in the SkyMapper catalog, denoted by the $\_petro$ flag.

Since each of the 32 CCD chips of the SkyMapper camera have slightly different sensitivities, we needed to apply additional photometric zeropoint corrections for data taken with each of the 64 readout amplifiers.
We derived these offsets by first applying the overall zeropoint offsets computed for each exposure and combining the source catalogs from each night.
We then repeated the procedure outlined in the previous paragraph, except we computed the residual zeropoint offset for sources on each of the 64 readout amplifiers.

For each night, our images were then stacked by sigma-clipping 5$\sigma$ outliers to remove cosmic rays after incorporating the above magnitude zeropoint corrections using the SWARP package \citep{bmr+02}.
Then, a final mosaic was generated by stacking the nightly images following the aforementioned procedure.
Final source catalogs were computed with Source Extractor package with the same configuration as used in deriving initial catalogs for the zeropoint calibration.
All the magnitudes reported in this paper are from the MAG\_AUTO keyword in Source Extractor, which are magnitudes obtained with elliptical apertures.
We opted not to use magnitudes derived from fitting a point spread function (PSF) of the images as the PSF appeared to not be stable over the full field of view of the SkyMapper images.
Our final source catalogs were de-reddening following \citet{wol+18}, based on reddening maps from \citet{sfs+98}.
An additional zero-point correction was applied to the $u$ and $v$ magnitudes following equations 4 and 5 in \citet{cwm+19} to account for a reddening overcorrection for sources close to the Galactic plane.

\subsubsection{Pattern-Noise Removal}
\label{sec:photpattern}

\begin{figure*}[!htbp]
\centering
\includegraphics[width =\textwidth]{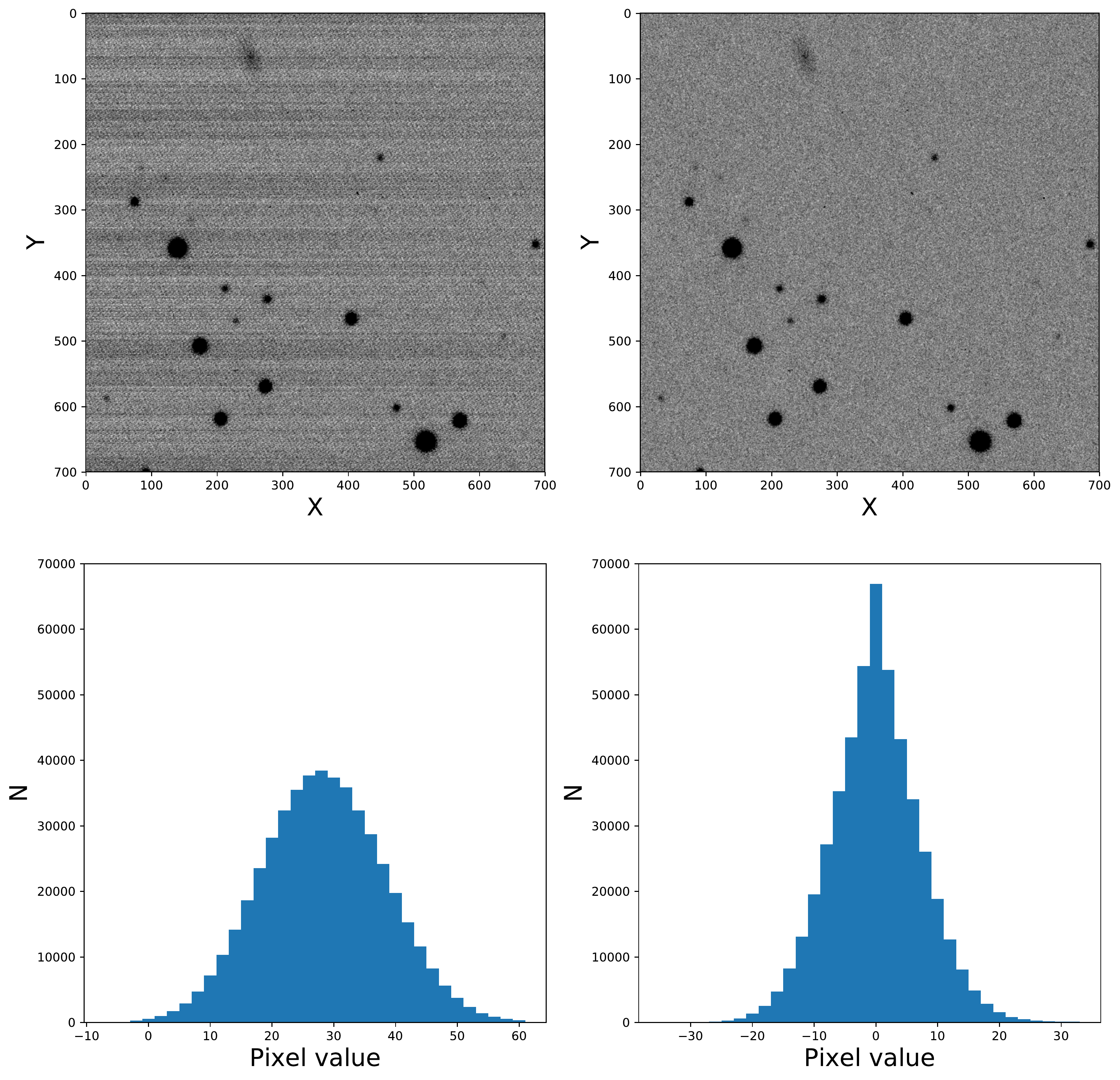}
\caption{Top: Comparison of a portion of an image immediately before (left) and after (right) pattern noise removal (see Section~\ref{sec:photpattern}). Bottom: A histogram of pixel values in each image before pattern noise removal (left) and after removal (right). After pattern-noise removal, the standard deviation of the values of the pixels in this image is $\sim$7 counts. This spread agrees with the range of readout noises reported for the SkyMapper readout amplifiers in \citet{wol+18}. }
\label{fig:patternnoise}
\end{figure*}

Before deriving the astrometric solution, we had to remove pattern-noise signatures that were imprinted on each image. 
This issue is illustrated in the left panel of Figure~\ref{fig:patternnoise}.
These signatures are composed of a high-frequency interference pattern and low-frequency waves with row-dependent zeropoint offsets.
These pattern-noise signatures changed between exposures and between each of the 64 readout amplifiers.

We performed the following steps to remove the pattern-noise after bias subtraction and flat-field correction.
To remove the low-frequency components of this signature, a 5th order polynomial was iteratively fit to each row of data from each readout amplifier after sigma-clipping outliers $2\sigma$ above the fit and $3\sigma$ below the fit.  
After this process, the high-frequency interference pattern was found to be identical across all CCD chips of a given image, after accounting for the orientation of each chip. 
Consequently, we aligned the orientation of each CCD chip and median-filtered the data from two sets of 32 amplifiers to obtain two templates of the high-frequency signature.
Each template was then fit, using a scaling factor and a zero-point offset as free parameters, to the appropriate readout amplifier and subtracted to remove the high-frequency pattern.
Figure~\ref{fig:patternnoise} shows a comparison of images before and after pattern noise removal.
Visually, the dominant systematic patterns appear to be removed.
Quantitatively, the standard deviation of the values of the background pixels drops by $\sim$30\% after our pattern noise removal procedure.

\subsubsection{Completeness \& Photometric Precision}
\label{sec:photcomplete}

We measure the completeness of our SkyMapper source catalog by comparing the number of stars in the DES Y1A1 gold catalog \citep{dsr+18} and in our catalog within $60^\prime$ of the center of Tucana II. 
We first convert the $g_{DES}$ magnitudes to SkyMapper $g_{\text{SM}}$ magnitudes using the following formula, which we derive by fitting a polynomial to our SkyMapper photometry using magnitudes from the DES Y1A1 gold catalog as independent variables: $g_{\text{SM}} = 0.983\times g_{\text{DES}} - 0.144\times(g_{\text{DES}} - i_{\text{DES}} ) + 0.431$.
We find that the cumulative 95\% completeness limits correspond to the following SkyMapper magnitude limits: $g_{\text{SM}} = 22.3$ for the SkyMapper $g$ filter, 
$g_{\text{SM}} = 22.1$ for the SkyMapper $i$ filter, $g_{\text{SM}} = 20.8$ for the SkyMapper $v$ filter, and $g_{\text{SM}} = 20.1$ for the SkyMapper $u$ filter.
We further find that the typical uncertainty in our photometry, as reported by \texttt{magerr\_auto} in Source Extractor, reaches 0.05\,mags at the following DES magnitudes:
$g_{\text{SM}} \sim 22.0$ for the SkyMapper $g$ filter, $g_{\text{SM}} \sim 21.6$ for the SkyMapper $i$ filter, $g_{\text{SM}} \sim 19.6$ for the SkyMapper $v$ filter, and $g_{\text{SM}} \sim 19.2$ for the SkyMapper $u$ filter.

As an additional check that our photometry is well-characterized, we compared the magnitudes derived from our pipeline to those in SkyMapper DR1.1 for stars in common between the two catalogs.
We find that the residuals between our magnitudes and those in SkyMapper DR1.1 are distributed as gaussians, therefore implying that our photometry is well-behaved with respect to the public SkyMapper data.
The standard deviation in the residuals reaches 0.05\,mags for $g_{\text{SM}}$ magnitudes when $g_{\text{SM}}\sim15.8$, $i_{\text{SM}}$ magnitudes when $i_{\text{SM}}\sim15.5$, $v_{\text{SM}}$ magnitudes when $v_{\text{SM}}\sim15.4$, and $u_{\text{SM}}$ magnitudes when $u_{\text{SM}}\sim15.8$.
We note that the scatter is driven by the uncertainty in the SkyMapper DR1.1 catalog, as our photometry is substantially deeper.

\begin{deluxetable}{lcr} 
\tablecolumns{3}
\tablewidth{\columnwidth}
\tablecaption{\label{tab:photobs} Photometric Observations}
\tablehead{   
  \colhead{Date} &
  \colhead{filter} & 
  \colhead{Exposure time (s)}
}
\startdata
2015 Jul 19 & $u$ & 24 $\times$ 300\\
2015 Jul 20 & $u$ & 21 $\times$ 300\\
2015 Jul 26 & $u$ & 9 $\times$ 300\\
2015 Aug 09 & $u$ & 20 $\times$ 300\\
2015 Aug 10 & $u$ & 15 $\times$ 300\\
2015 Jul 19 & $v$ & 24 $\times$ 300\\
2015 Jul 20 & $v$ & 20 $\times$ 300\\
2015 Jul 26 & $v$ & 5 $\times$ 300\\
2015 Aug 09 & $v$ & 22 $\times$ 300\\
2015 Aug 10 & $v$ & 16 $\times$ 300\\
2015 Aug 09 & $g$ & 4 $\times$ 300\\
2015 Dec 02 & $g$ & 4 $\times$ 300\\
2015 Dec 03 & $g$ & 4 $\times$ 300\\
2015 Dec 07 & $g$ & 4 $\times$ 300\\
2015 Dec 13 & $g$ & 4 $\times$ 300\\
2015 Dec 14 & $g$ & 4 $\times$ 300\\
2015 Dec 02 & $i$ & 8 $\times$ 300
\enddata
\end{deluxetable}

\section{Grid of synthetic photometry}
\label{sec:grid}

We generated a grid of flux-calibrated, synthetic spectra over a range of stellar parameters and metallicities specifically covering that expected for red giant branch (RGB) and main sequence turn off (MSTO) stars in Tucana II and the Milky Way halo.
The stellar parameters of this grid are given in Table~\ref{tab:grid}.
We then computed the expected flux through each of the SkyMapper filters from these synthetic spectra.


\begin{deluxetable}{cccc} 
\tablecolumns{3}
\tablewidth{\columnwidth}
\tablecaption{\label{tab:grid} Stellar parameters of synthetic spectrum grids}
\tablehead{   
  \colhead{Parameter} &
  \colhead{Minimum} &
  \colhead{Maximum} &
  \colhead{Step}
}
\startdata
\multicolumn{4}{c}{RGB Grid} \\
\hline
$\lambda$ & 3000\,\AA & 9000\,\AA & 0.01\,\AA\\
$T_{\text{eff}}$ & 4000\,K & 5700\,K & 100\,K\\
log $g$ & 1.0 & 3.0 & 0.5\\
$[$Fe/H$]$ & $-$4.0 & $-$0.5 & 0.5\\
\hline
\multicolumn{4}{c}{MSTO Grid} \\
\hline
$\lambda$ & 3000\,\AA & 9000\,\AA & 0.01\,\AA\\
$T_{\text{eff}}$ & 5600\,K & 6700\,K & 100\,K\\
log $g$ & 3.0 & 5.0 & 0.5\\
$[$Fe/H$]$ & $-$4.0 & $-$0.5 & 0.5\\
\enddata

\end{deluxetable}


\subsection{Generating synthetic spectra}
\label{sec:synspec}

\begin{figure*}[!htbp]
\centering
\includegraphics[width =\textwidth]{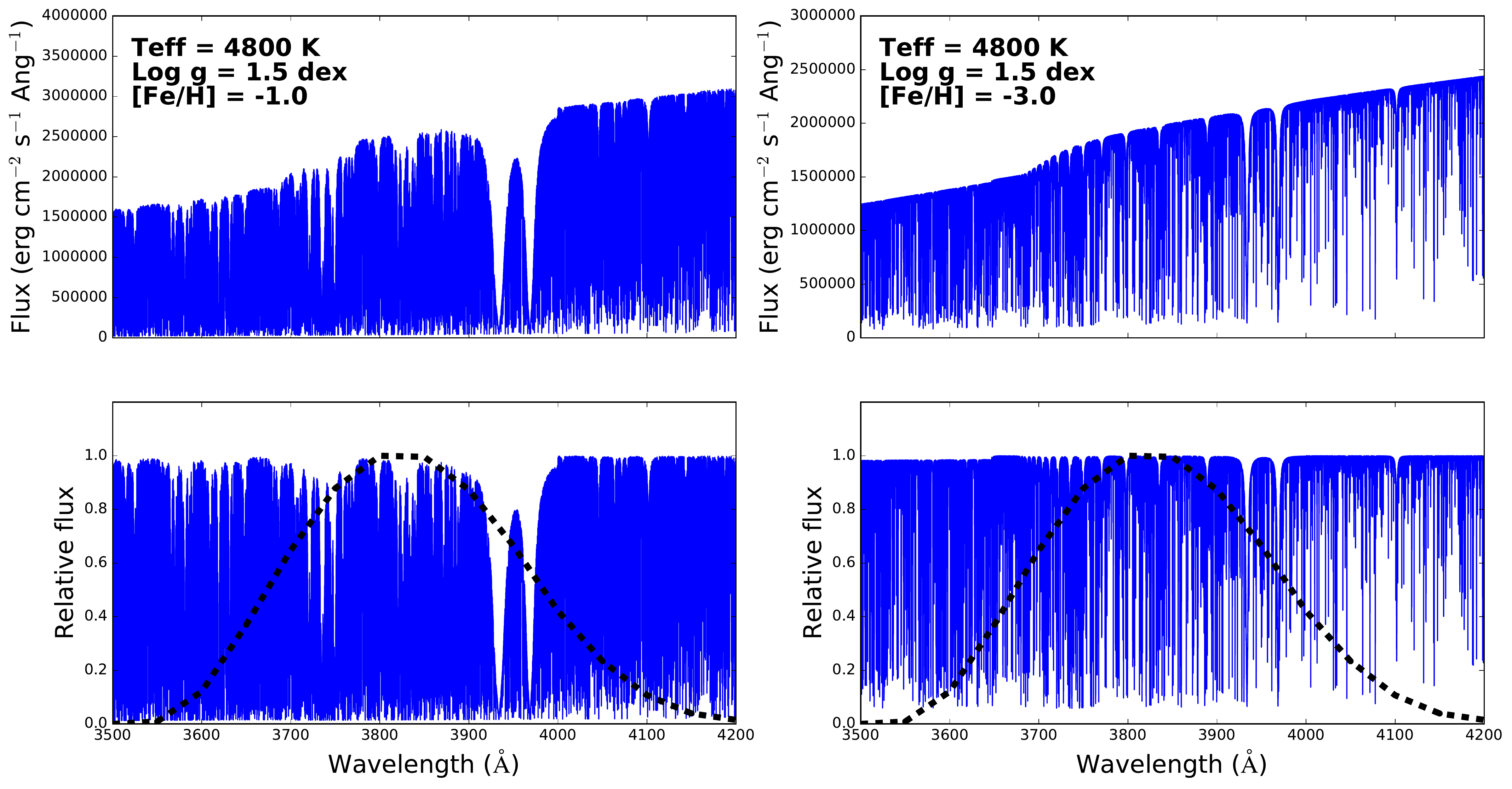}
\caption{Top: Flux-calibrated synthetic spectra for two stars with the same stellar parameters, but different metallicities. Bottom: Normalized synthetic spectra for the same two stars, with the bandpass of the SkyMapper $v$ filter overplotted as a dashed line. 
The strength of the Ca II K line measurably affects the flux through the $v$ filter. 
The CN absorption feature at $3870$\,{\AA} is also sufficiently prominent to affect the flux through the filter, and its impact on the measured photometric metallicity is discussed in Section~\ref{sec:carbon}.}
\label{fig:synthspec}
\end{figure*}

We used the Turbospectrum synthesis code \citep{ar+98, p+12}, MARCS model atmospheres \citep{gee+08}, and a linelist composed of all lines between 3000\,{\AA} to 9000\,{\AA} available in the VALD database \citep{pkr+95, rpk+15} to generate our grid of flux-calibrated synthetic spectra.
We replaced the lines of the CN molecule in the VALD line list with those from \citet{brw+14} and \citet{slr+14}, those of CH with lines from \citet{mpv+14} and \citet{bbs+13}, and those of C$_2$ with lines from \citet{rbb+14}.
This resulted in a linelist with $\sim 800,000$ lines.
An example of two synthetic spectra with different metallicities is shown in Figure~\ref{fig:synthspec}.
The ${}^{12}$C/${}^{13}$C isotope ratio was assumed following the relation presented in \citet{kgz+15}, which is based on figure~4 in \citet{kps+01}.
We note that solar abundances for the MARCS model atmospheres are adopted from \citet{gas+07}.

For the analysis of RGB stars, we generated a grid of spectra with stellar parameters ranging from $T_{\text{eff}}$ = 4000\,K to 5700\,K, $\log g$ = 1.0 to 3.0, and [Fe/H] = $-4.0$ to $-0.5$.
We opted to use the ``standard" spherical model geometry within the MARCS model atmospheres. 
We used a microturbulence of 2\,km\,s$^{-1}$.
The $\alpha$-enhancement was set to [$\alpha$/Fe] = 0.4 for stars with [Fe/H] $< -1.0$, and linearly decreased between $-1 < \mbox{[Fe/H]} < 0$ such that [$\alpha$/Fe] = 0 when [Fe/H] = 0.

For the analysis of MSTO stars, we generated a grid of spectra with stellar parameters ranging from $T_{\text{eff}}$ = 5600\,K to 6700\,K, $\log g$ = 3.0 to 5.0, and [Fe/H] = $-4.0$ to $-0.5$.
We opted to use plane-parallel model geometries as part of the MARCS model atmospheres, and used the same microturbulence value and [$\alpha$/Fe] trends as for the RGB grid.

\subsection{Generating synthetic photometry}
\label{sec:synphot}

For each synthetic spectrum, we calculated the absolute magnitude from the flux through each of the SkyMapper $u$, $v$, $g$, and $i$ filters.
First, we retrieved the bandpasses of each of the SkyMapper filters \citep{bbs+11} from the Spanish Virtual Observatory (SVO) Filter Profile Service \citep{rsb+12}\footnote{http://svo2.cab.inta-csic.es/svo/theory/fps3/}.
We then closely followed the methodology in \citet{cv+14} to generate synthetic magnitudes for each of our synthetic spectra. 
We computed synthetic \texttt{AB} magnitudes, in which a flux density of $F_{\nu} = 3.63 \times 10^{-20}\,\text{erg}\,\text{cm}^{-2}\,\text{s}^{-1}\,\text{Hz}^{-1}$ is defined as having m$_{\texttt{AB}} = 0$, through each filter by applying the following formula

\begin{equation}
m_{\text{AB}} = -2.5\log\frac{\int_{\nu_i}^{\nu_f}\nu^{-1}\,F_{\nu}\,T_{\eta}\,d\nu}{\int_{\nu_i}^{\nu_f}\nu^{-1}\,T_{\eta}\,d\nu} - 48.60
\label{eqn:ACF}
\end{equation}

\noindent where $F_\nu$ is the flux from a flux-calibrated synthetic spectrum as a function of wavelength, $T_{\eta}$ is the system response function over the bandpass of the filter, and $\nu_i$ and $\nu_f$ are the lowest and highest wavelengths of the bandpass filter.
We note that small zero point offsets on the order of $\sim0.02$ are possible in this formalism, as briefly mentioned in \citet{cv+14}.
We, however, find that we need to apply an a zero-point correction of +0.06\,mags to our synthetic $v$ magnitudes to derive accurate metallicities, as described in the second to last paragraph of Section~\ref{sec:GCs}.
\\
\begin{figure*}[!htbp]
\centering
\includegraphics[width =\textwidth]{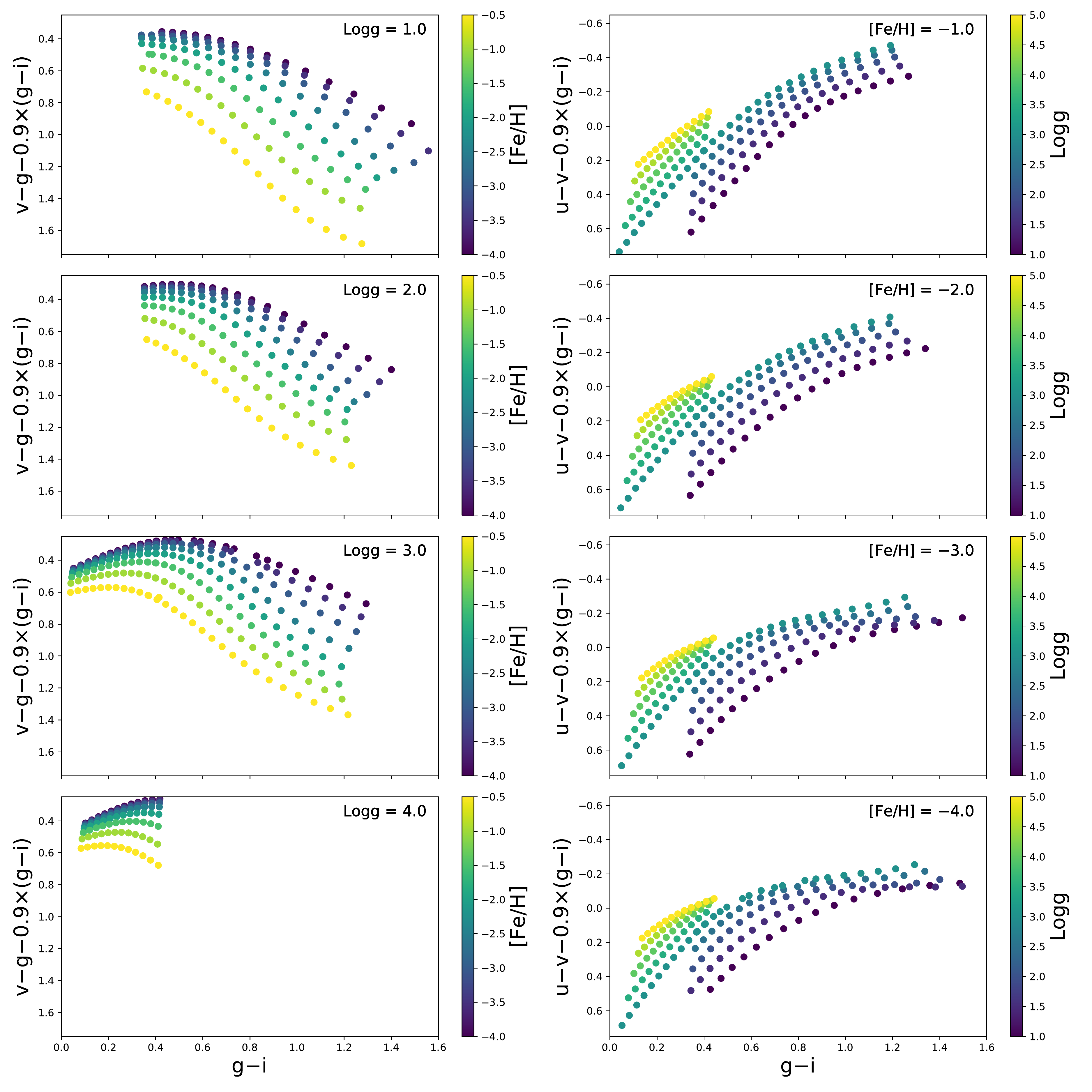}
\caption{Left: Contours used to measure photometric metallicities from the SkyMapper photometry for several $\log\,g$ values. Right: Contours used to measure photometric surface gravities from SkyMapper photometry for several [Fe/H] values.}
\label{fig:synthfeh}
\end{figure*}

\section{Analysis}
\label{sec:analysis}

One aim of this study is to demonstrate that SkyMapper photometry can be used to derive stellar parameters and metallicities to ultimately derive a metallicity distribution function for the Tucana II UFD.
For that purpose, it is necessary to carefully characterize the sources of uncertainty in our photometric metallicities.
Therefore, we first describe our methodology in measuring photometric metallicities and surface gravities in Sections~\ref{sec:metallicities} and~\ref{sec:loggs}. 
Then, in Sections~\ref{sec:carbon} through Section~\ref{sec:medres}, we attempt to quantify the impact of (strong) carbon molecular features within the bandpass of the $v$ filter on our photometric metallicities, and compare our photometric metallicities to available spectroscopic metallicities in the literature.

\subsection{Measuring metallicities from photometry}
\label{sec:metallicities}

The SkyMapper $v$ filter has been designed to be sensitive to stellar metallicity \citep[e.g.,][]{ksb+07}.
This sensitivity arises from the presence of the strongest metal absorption line, Ca II K, at 3933.7{\,\AA} in the bandpass.
Since the strength of the Ca II K line scales with the overall metallicity of the star, the overall flux measured through the filter is thus governed by the stellar metallicity.
An example is illustrated in Figure~\ref{fig:synthspec}, where synthetic spectra of stars with [Fe/H] = $-1.0$ and $-3.0$ are juxtaposed, and the bandpass of the $v$ filter is overplotted.

Making use of the relation between Ca II K absorption and metallicity, previous work by  \citet{ksa+12} suggested that metal-poor stars can be discriminated from metal-rich ones in the $v-g-2\times(g-i)$ vs. $g-i$ space. 
Inspired by this, we instead choose to utilize $v-g-0.9\times(g-i)$ vs. $g-i$ as a discriminator, which we have already successfully used to identify metal-poor dwarf galaxy stars using our custom SkyMapper data and SkyMapper DR1.1 \citep{cfj+18, cf+19}.

As part of this work, we plotted $v-g-0.9\times(g-i)$ vs. $g-i$ of the photometry from the synthetic spectra (described in Section~\ref{sec:grid}) on the left panels of Figure~\ref{fig:synthfeh}. We did so for four different $\log g$ values, from 1 to 4. 
As can be seen, stars of a given metallicity form well-behaved contours allowing us to easily interpolate between these contours to derive quantifiable stellar metallicities.
Hence, we interpolated between these metallicity contours with a piecewise 2d cubic spline interpolator using the \texttt{scipy.interpolate.griddata} function, and thereby derived photometric metallicities for every star with $v$, $g$, and $i$ photometry.
We flagged each star with photometry placing them beyond the upper (most metal-poor) bounds given of these contours, and set their metallicity to the boundary value (i.e., $\mbox{[Fe/H]}=-4.0$).

As is shown on the left panel in Figure~\ref{fig:synthfeh},  the contours used for measuring metallicity depend on the surface gravity $\log g$. It is thus necessary to assume an initial $\log g$ before attempting any metallicity calculations. We initually assume $\log g = 2$  which should roughly correspond to the surface gravity of stars on the RGB of Tucana II.

Upon obtaining these initial metallicities, photometric surface gravities were then derived as described in Section~\ref{sec:loggs}. After that, the photometric surface gravities were used to determine our final photometric metallicities.
The average change in the photometric metallicities upon updating the surface gravity is marginal ($\sim0.02$\,dex for stars with $\log g$ $<$ 3.0), suggesting that one iteration is sufficient for convergence.
Section~\ref{sec:sigmas} discusses our final adopted metallicity uncertainties.

\subsection{Measuring surface gravities from photometry}
\label{sec:loggs}

\begin{figure*}[!htbp]
\centering
\includegraphics[width =\textwidth]{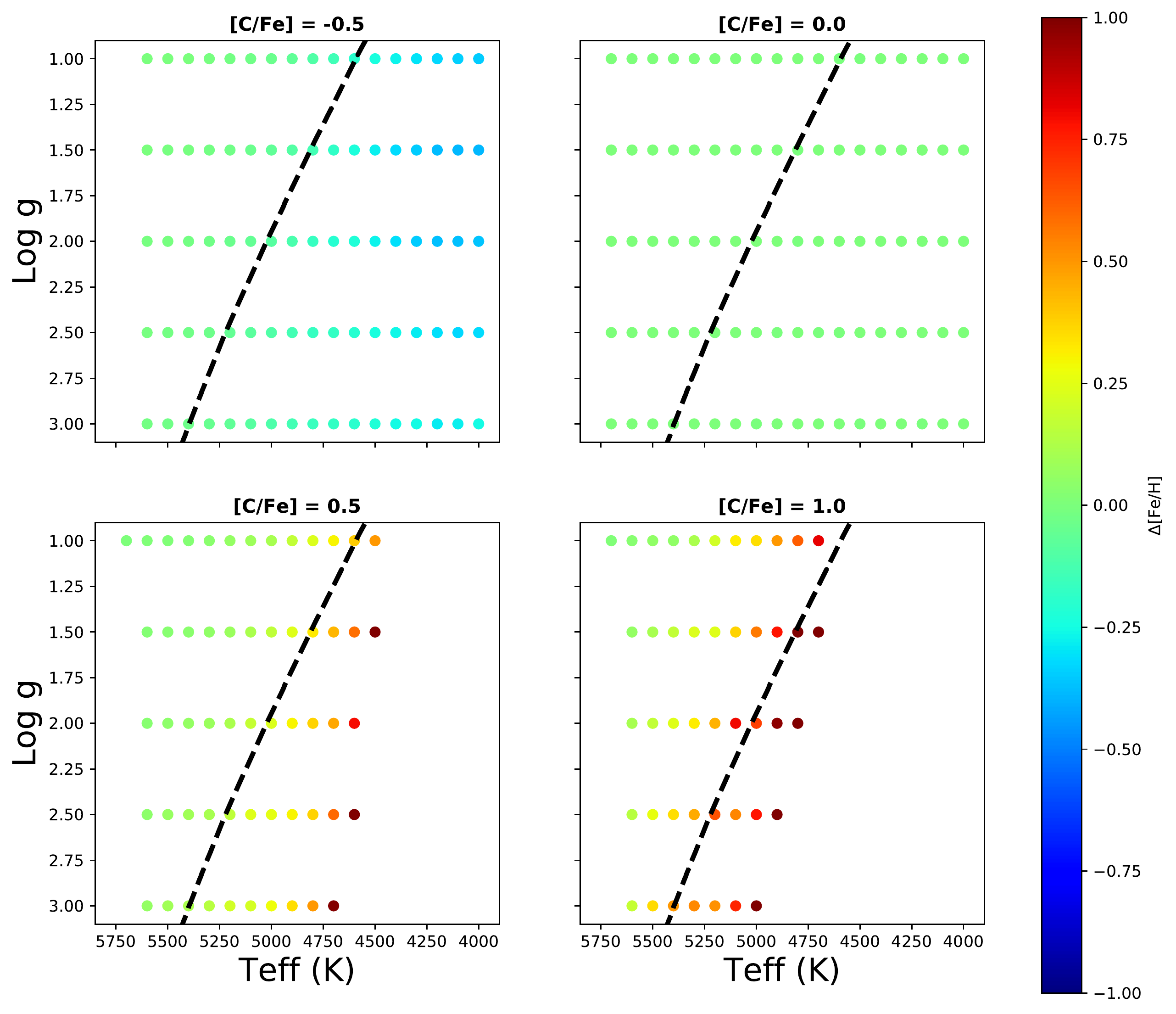}
\caption{Effect of different carbon abundances on the photometric metallicities as a function of surface gravity and effective temperature and a fixed [Fe/H] = $-2.5$.
A strong effect occurs towards lower effective temperatures ($\lesssim$4700\,K) and with increasing carbon enhancement ([C/Fe] $\ge$ 0.5) because the strength of the CN feature at $\sim$3870\,{\AA} has a significant effect on the flux through the $v$ filter.
The dashed line corresponds to the RGB of a [Fe/H] =$-2.5$, 12\,Gyr Dartmouth isochrone \citep{dcj+08}.}
\label{fig:Carbon}
\end{figure*}

The SkyMapper $u$ and $v$ filters can be used to discriminate the surface gravities of stars. These filters  bracket the Balmer Jump at 3646\,{\AA} which is sensitive to the H$^{-}$ opacity, which, in turn, is a function of the $\log g$ of the star \citep[e.g.,][]{mks+11}. Similar to the method outlined in Section~\ref{sec:metallicities} for deriving photometric metallicities, plotting $u-v-0.9\times(g-i)$ vs. $g-i$ can discriminate stellar surface gravities.
This behavior is illustrated on the right panels in Figure~\ref{fig:synthfeh}, based on the synthetic spectra described in Section~\ref{sec:grid}. We then use the same interpolation technique described in Section~\ref{sec:metallicities} to derive photometric surface gravities from the $u-v-0.9\times(g-i)$ contours.

We use the first-pass synthetic metallicities from Section~\ref{sec:metallicities} to choose the corresponding set of $\log g$ contours from which to derive surface gravities. 
Since we  use surface gravities solely to remove foreground main-sequence stars from our sample, we opt not to iteratively re-measure $\log\,g$ with updated photometric metallicity values.

\subsection{Dependence of the photometric metallicity measurements on carbon abundance}
\label{sec:carbon}

\begin{figure*}[!hbt]
\centering
\includegraphics[width =\textwidth]{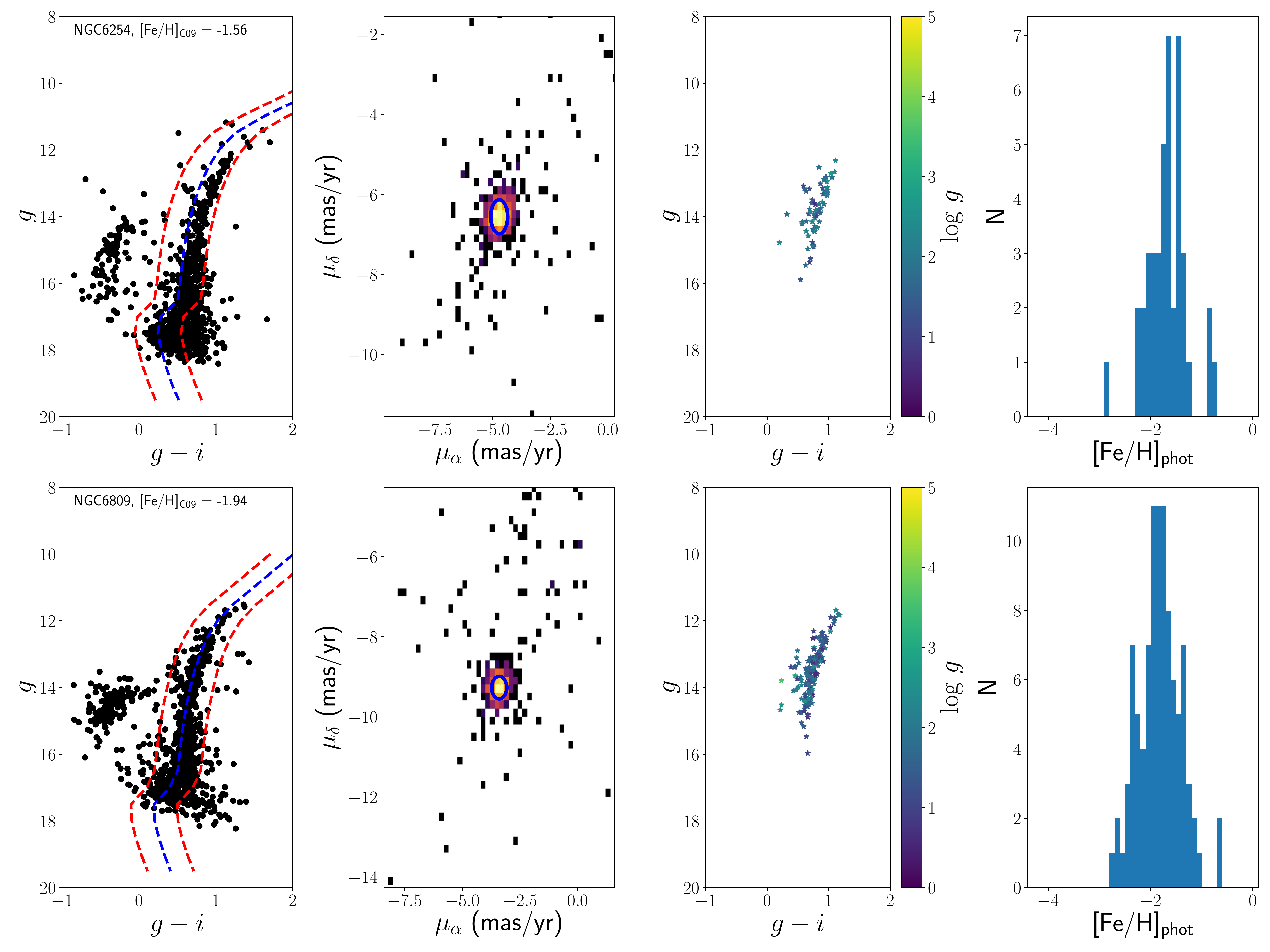}
\caption{Overview of the process of finding likely member stars to derive mean photometric metallicities for NGC6254 (top), a cluster with [Fe/H] = $-1.56$, and NGC6809 (bottom), a cluster with [Fe/H] = $-1.94$ \citep{cbg+09}. 
From left to right: selection of candidate members using a Dartmouth isochrone \citep{dcj+08}; further selection using \textit{Gaia} proper motion data: a 2d Gaussian (blue) is fitted to a density plot of the proper motion data to select likely members; resulting CMD after only retaining stars with photometric $\log\,g < 3$ and with proper motions in the fitted 2d Gaussian; histogram of photometric metallicities of the stars that passed all selection criteria.}
\label{fig:GCselection}
\end{figure*}

Since a CN molecular absorption feature is located at $\sim$3870\,{\AA} in the bandbass of the SkyMapper $v$ filter (encompassing $\sim3600$\,{\AA} to $\sim4100$\AA), its spectral morphology can become strong enough to systematically affect the flux through the filter. The strength of the CN feature significantly depends on the carbon abundance and effective temperature of the star: A higher carbon abundance and a lower effective temperature  leads to a stronger CN absorption feature which, in turn, leads to an artificially higher photometric metallicity. 
This could systematically skew photometric metallicity results since a significant fraction of metal-poor stars tend to be enhanced in carbon ([C/Fe] $> 0.7$) and are known as carbon-enhanced metal-poor (CEMP) stars. 
80\% of stars with [Fe/H] $< -4.0$ and still 24\% of stars with [Fe/H] $< -2.5$ have [C/Fe] $> 0.7$  \citep{pfb+14}.
We thus attempt to quantify these effects to gauge how our photometric metallicities derived from the $v$ filter are influenced by the carbon abundance. 

We test the effect of the strength of the CN feature on the flux through the SkyMapper $v$ filter by regenerating our grid of synthetic spectra, as described in Section~\ref{sec:grid}. 
We do so by varying the carbon abundances of the synthetic spectra between [C/Fe] = $-0.5$ and [C/Fe] = 1.0 in intervals of 0.5.
We then derived the $v - g - 0.9\times (g-i)$ index and $g-i$ colors for these synthetic spectra and obtained the corresponding photometric metallicities following the procedure described in Section~\ref{sec:metallicities}.

In Figure~\ref{fig:Carbon}, we plot the changes in the resulting photometric metallicity, relative to a baseline of [C/Fe] = 0, for four different carbon abundances as a function of surface gravity and effective temperature, corresponding to our RGB grid in Table~\ref{tab:grid}. 
The [C/Fe] = 0 baseline, however, is temperature-dependent as the strength of the CN band is temperature sensitive. Decreasing the carbon abundance by 0.5\,dex to gauge the corresponding effect on the photometric metallicity leads minimal effects of $\sim 0.1$\,dex, as seen in Figure~\ref{fig:Carbon}. 
But among the cooler stars ($T_{\text{eff}} < 4700$\,K) with [C/Fe] $> 0.5$, we find significant changes (the ultimately measured [Fe/H] of a CEMP star gets artificially increased by $\Delta$[Fe/H]$ \sim 0.5$). 
Accordingly, some true CEMP stars may always remain ``hidden" in our samples, especially among the cooler stars.
No significant effects are apparent otherwise since the CN feature is relatively weak for all these stars, meaning the feature does not influence the overall flux through the $v$ filter.
Therefore, our analysis suggests that we would still select moderately cool ($T_{\text{eff}} \sim 4700$\,K) CEMP stars as very metal-poor candidates.
This result is supported by the fact that we can re-identify all three members of Tucana II that are CEMP stars, all of which are warmer than 4600\,K \citep{cfj+18} but are not greatly enhanced in carbon ([C/Fe] $< 1.0$ before applying the carbon correction following \citet{pfb+14}).

Overall, we conclude that while our derived photometric metallicities are dependent on the carbon abundance and effective temperature of stars, we can select stars with (uncorrected) $\mbox{[C/Fe]}\lesssim0.5$ that also have [Fe/H] $\sim -2.5$ and have $T_{\text{eff}} \gtrsim 4800$\,K with sufficient precision (within +0.3\,dex). 
Stars which turn out to have e.g.,  [C/Fe] $\sim$ 0.35, if on the RGB ($\log\,g \sim 1.5$), would likely become CEMP stars after applying a correction for the evolutionary status of the star following \citet{pfb+14}. 
Correspondingly, a star with a carbon abundance of up to [C/Fe] $\sim 1$ and [Fe/H] $\sim -2.5$ could principally still be identified as a candidate member of a UFD ($\mbox{[Fe/H]} \lesssim -1.5)$ when it has $T_{\text{eff}} \gtrsim 5000$\,K since its photometric metallicity would be shifted by $\Delta$[Fe/H]$\lesssim1.0$.

CEMP-s stars tend to be the most carbon-enhanced ([C/Fe] $\gtrsim$ 1.25 at $\mbox{[Fe/H]}\gtrsim-2.8$) subclass of CEMP stars and would thus systematically have among the strongest CN bands \citep{ybp+16} somewhat irrespective of temperature.
Consequently, CEMP-s stars would appear as much more metal-rich stars (we estimate by about 1\,dex or more) supposing they were on the RGB. 
Hence, any CEMP-s star with e.g., [Fe/H] = $-2.5$ would be measured as having a metallcity of $-$1.5\,dex or higher. 
These metallicities tend to be excluded from our selection since we are focused on more metal-poor stars, so our sample could thus be regarded mostly free of CEMP-s stars.

At lower metallicities, the effect of the CN absorption on the measured [Fe/H] is somewhat mitigated.
After repeating our procedure for [Fe/H] $= -4.0$, a star with [C/Fe] $= 1.0$ will affect the measured [Fe/H] by $\sim$ 0.3\,dex when it has $T_{\text{eff}} \gtrsim 4500$\,K. 
When [C/Fe] is increased to 2.0, moderately cool ($T_{\text{eff}} \sim 4800$\,K) stars will be affected by up to 0.75\,dex due to the highly nonlinear growth of the CN feature. 
Nevertheless, we note that to first order, the CN feature of a star with [Fe/H] = $-5.0$ and [C/Fe] = $2.0$ should be similar to that of a star with [Fe/H] = $-4.0$ and [C/Fe] = 1.0, assuming similar stellar parameters.
This implies that most CEMP stars with $\mbox{[Fe/H]}<-4.0$ should be identifiable, except for perhaps very extreme cases.

In general, these results suggests that any metallicity distribution function (MDF) derived from SkyMapper photometry will likely be upscattered to some degree.
According to our models, the most metal-poor ([Fe/H] $\sim -4.0$), moderately cool ($T_{\text{eff}} \sim 4800$\,K) stars with a significant carbon-enhancement ([C/Fe] $\sim 2.0$) will be affected by up to 0.75\,dex. 
High-resolution spectroscopic follow-up observations of stars in the SkyMapper dataset can confirm the extent of these CN feature-induced changes.

\begin{figure}[!htbp]
\centering
\includegraphics[width =\columnwidth]{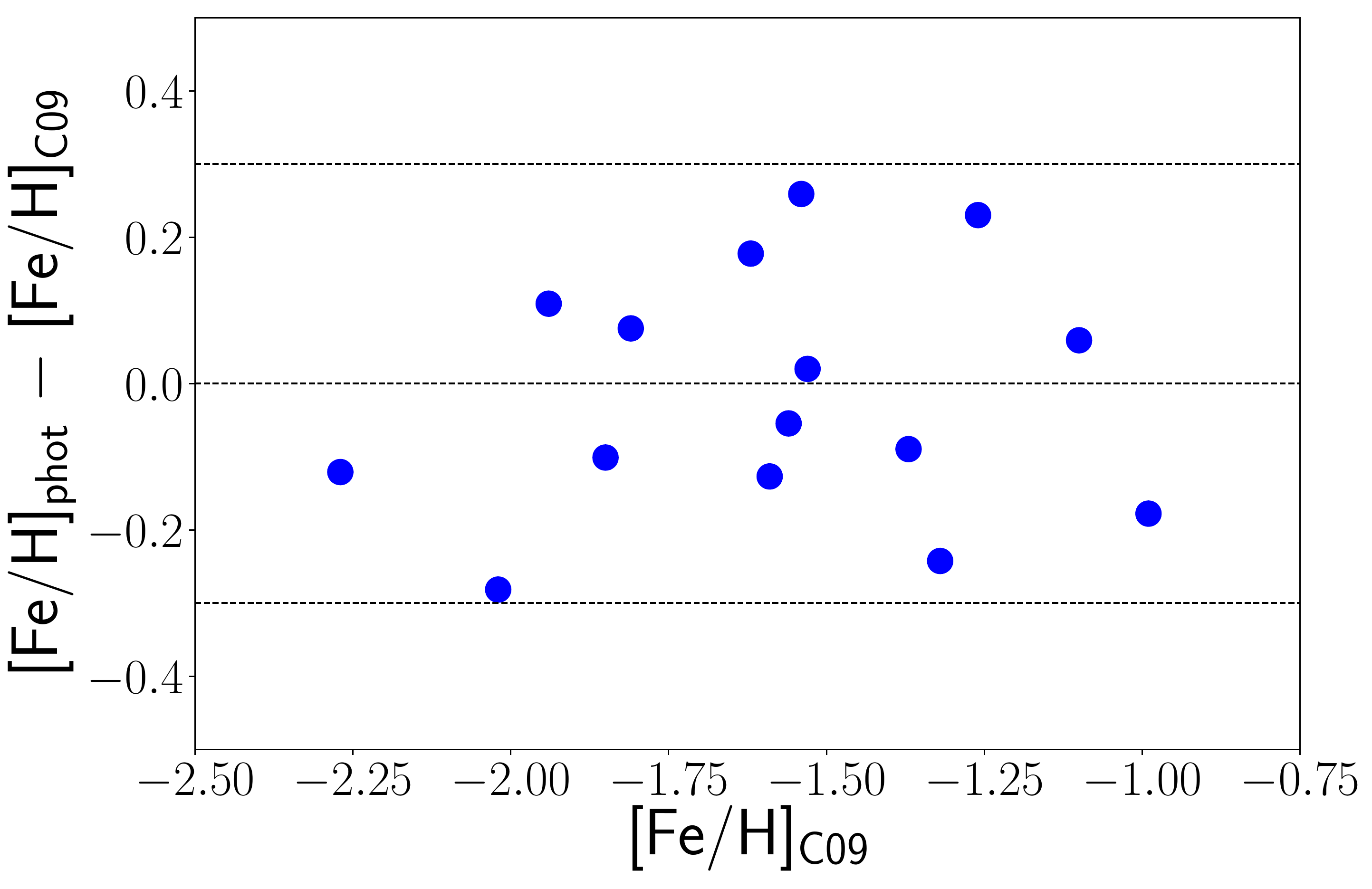}
\caption{Residuals of our photometric metallicities of globular clusters with respect to spectroscopic results from \citet{cbg+09}, as a function of their metallicities. These residuals have a standard deviation of 0.16\,dex. The standard error in the mean of the residuals is $\sigma_{\Delta \text{[Fe/H]}} = 0.04$.}
\label{fig:GCmetallicities}
\end{figure}

\subsection{Comparison to Globular Clusters}
\label{sec:GCs}

Globular clusters are old ($\sim10$\,Gyr) and metal-poor ([Fe/H] $< -1.0$) star clusters. 
The dispersion of the metallicities of their member stars is generally small ($\sigma_{\text{[Fe/H]}} \sim 0.05$) as discussed in e.g., \citet{cbg+09}.
The stellar population of individual globular clusters therefore provides a useful test of the precision of deriving photometric metallicities, since most of the dispersion in the photometric metallicities of member stars can be ascribed to the uncertainty in our SkyMapper data and broader methodology.

While we did not observe any globular clusters as part of our observing program, a number of globular clusters are located within the footprint of the first data release (DR1.1) of the SkyMapper Southern Survey \citep{wol+18}.
Thus, we retrieved SkyMapper $u,v,g,$ and $i$ photometry from the DR1.1 catalog for member stars of all globular clusters in the southern hemisphere with a distance less than $10$\,kpc.
Specifically, we used the \citet{h+10} catalog of globular clusters, an update to the older \citet{h+96} catalog, and retrieved all stars within three times the tidal radius of each globular cluster.
This resulted in the retrieval of 17 globular clusters with metallicities ranging from [Fe/H] = $-2.27$ to [Fe/H] $= -0.99$ as measured in \citet{cbg+09}.

Upon retrieving SkyMapper $u,v,g,i$ photometry from the public catalog, we selected likely member stars of each globular cluster.
We first overlaid 13\,Gyr Dartmouth isochrones \citep{dcj+08} with metallicities and distance moduli matching those from \citet{cbg+09} and then selected all stars with $g-i$ within 0.3\,mag of the isochrone.
We then computed photometric metallicities and surface gravities of these candidate members using the methods described in Sections~\ref{sec:metallicities} and~\ref{sec:loggs}.

We used proper motion measurements from the second data release from the {\it Gaia} mission \citep[DR2;][]{gaia+16, gaia+18, sgg+17} as an additional avenue to exclude non-members since members of a globular cluster should have similar proper motions.
To identify the systemic proper motion of each cluster, we generated 2d histograms of the proper motions with a binsize of 0.2\,mas/yr in $\mu_\alpha$ and $\mu_\delta$. 
We then fitted a 2d elliptical Gaussian to the overdensity in each proper motion histogram and selected all stars enclosed within the 3$\sigma$ bounds of the Gaussian.

We then chose to only retain stars with photometric surface gravities of $\log\,g < 3$, as the depth of the public SkyMapper photometry with usable photometric metallicity precision ($g\sim16$) does not extend to the main sequences in our sample of globular clusters.
We also only retained stars with photometric metallicity uncertainties below 0.5\,dex and photometric surface gravity uncertainties below 0.75\,dex.
The determination of the uncertainties is described in Section~\ref{sec:sigmas}.
Each step of our selection of likely member stars of globular clusters is shown in Figure~\ref{fig:GCselection}. 
We finally derive an overall metallicity of each globular cluster by taking the average of the photometric metallicities, weighted by the inverse-squared photometric metallicity uncertainties, of each remaining sample of likely member stars.
Two histograms of the photometric metallicities of likely member stars for NGC6254 and NGC6809 are shown as examples in Figure~\ref{fig:GCselection}.

For all clusters for which we identify $N > 1$ member stars, we plotted the residuals of our derived cluster metallicities with respect to spectroscopically-derived metallicities of each cluster from \citet{cbg+09}.
We found that applying an offset of 0.06\,mags to our synthetic $v$ magnitudes removed a $\sim0.1$\,dex systematic offset between our metallicities and those in \citet{cbg+09}.
Thus, we applied this offset to our contours and re-derived our cluster metallicities.
We note that all photometric metallicities and $\log\,g$ values presented in this paper are calculated with this +0.06\,mags offset in the synthetic $v$ magnitudes.

The final residuals of our cluster metallicities with respect to \citet{cbg+09} are shown in Figure~\ref{fig:GCmetallicities}.
A negligible final offset of $-0.02$\,dex is found with a standard deviation is 0.16\,dex with respect to the values from \citet{cbg+09}. 
We thus take 0.16\,dex  as an estimate of the intrinsic uncertainty from our method as further discussed in Section~\ref{sec:sigmas}. 
Table~\ref{tab:GCs} lists our measured photometric metallicities as well as spectroscopic metallicities from \citet{cbg+09}.


\begin{deluxetable}{ccccc} 
\tablecolumns{5}
\tablewidth{\columnwidth}
\tablecaption{\label{tab:GCs} Photometric metallicities of globular clusters based on SkyMapper DR1.1 data}
\tablehead{   
  \colhead{Name} &
  \colhead{$(m-M)$} &
  \colhead{[Fe/H]$_{\text{C09}}$} &
  \colhead{[Fe/H]$_{\text{phot}}$} &
  \colhead{N}
}
\startdata
NGC7099 & 14.54 & $-$2.27 & $-$2.39 & 10\\
NGC6397 & 11.81 & $-$2.02 & $-$2.30 & 29\\
NGC6809 & 13.66 & $-$1.94 & $-$1.83 & 97\\
NGC4833 & 14.10 & $-$1.85 & $-$1.95 & 3\\
NGC6541 & 14.38 & $-$1.81 & $-$1.73 & 7\\
NGC6681 & 14.77 & $-$1.62 & $-$1.44 & 2\\
NGC3201 & 13.45 & $-$1.59 & $-$1.72 & 23\\
NGC6254 & 13.22 & $-$1.56 & $-$1.61 & 42\\
NGC6752 & 13.01 & $-$1.54 & $-$1.28 & 68\\
NGC5139 & 13.58 & $-$1.53 & $-$1.51 & 185\\
NGC6218 & 13.41 & $-$1.37 & $-$1.46 & 28\\
NGC288 & 14.75 & $-$1.32 & $-$1.56 & 7\\
NGC362 & 14.67 & $-$1.26 & $-$1.03 & 15\\
NGC6723 & 14.70 & $-$1.10 & $-$1.04 & 7\\
NGC6362 & 14.40 & $-$0.99 & $-$1.17 & 4
\enddata
\end{deluxetable}


\begin{figure*}[!htbp]
\centering
\includegraphics[width =\textwidth]{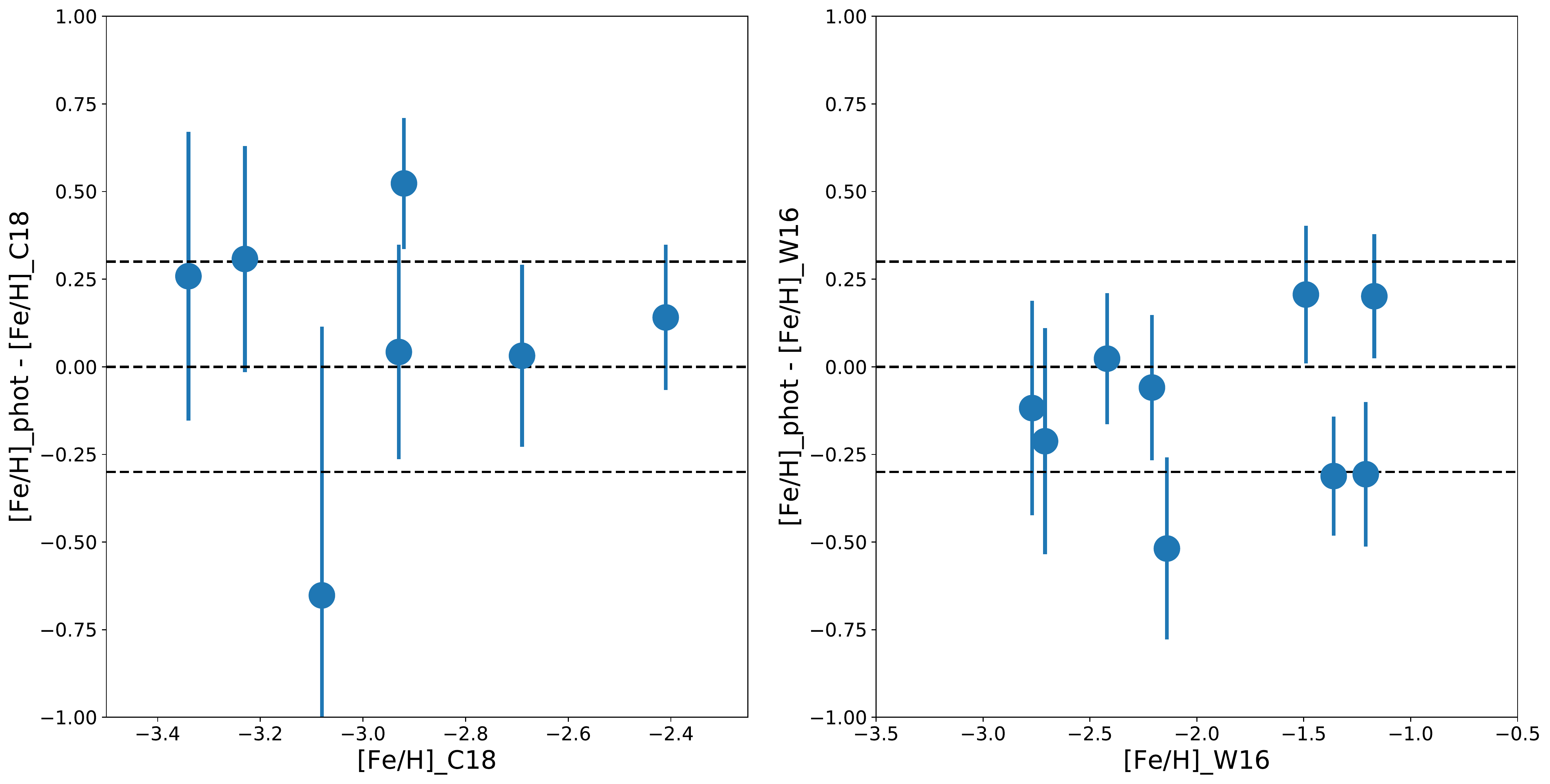}
\caption{Left: Comparison of our photometric metallicities to those derived from high-resolution spectroscopy in \citet{cfj+18}. Right: Comparison of our photometric metallicities to those derived from medium-resolution spectroscopy in \citet{wmo+16} for all stars in that paper with metallicity values with an uncertainty less than 0.2\,dex and [Fe/H] $< -1.0$.
We note that \citet{wmo+16} applied a zero-point offsets of either 0.16\,dex or 0.32\,dex to their metallicities that may account for the zero-point offset between our measurements and those in \citet{wmo+16}.
Dashed lines are drawn at $+0.3$ and $-0.3$ to guide the eye.
Error bars correspond to the uncertainty in the photometric metallicities of these stars.}
\label{fig:feh_compare}
\end{figure*}

\begin{figure}[!htbp]
\centering
\includegraphics[width =\columnwidth]{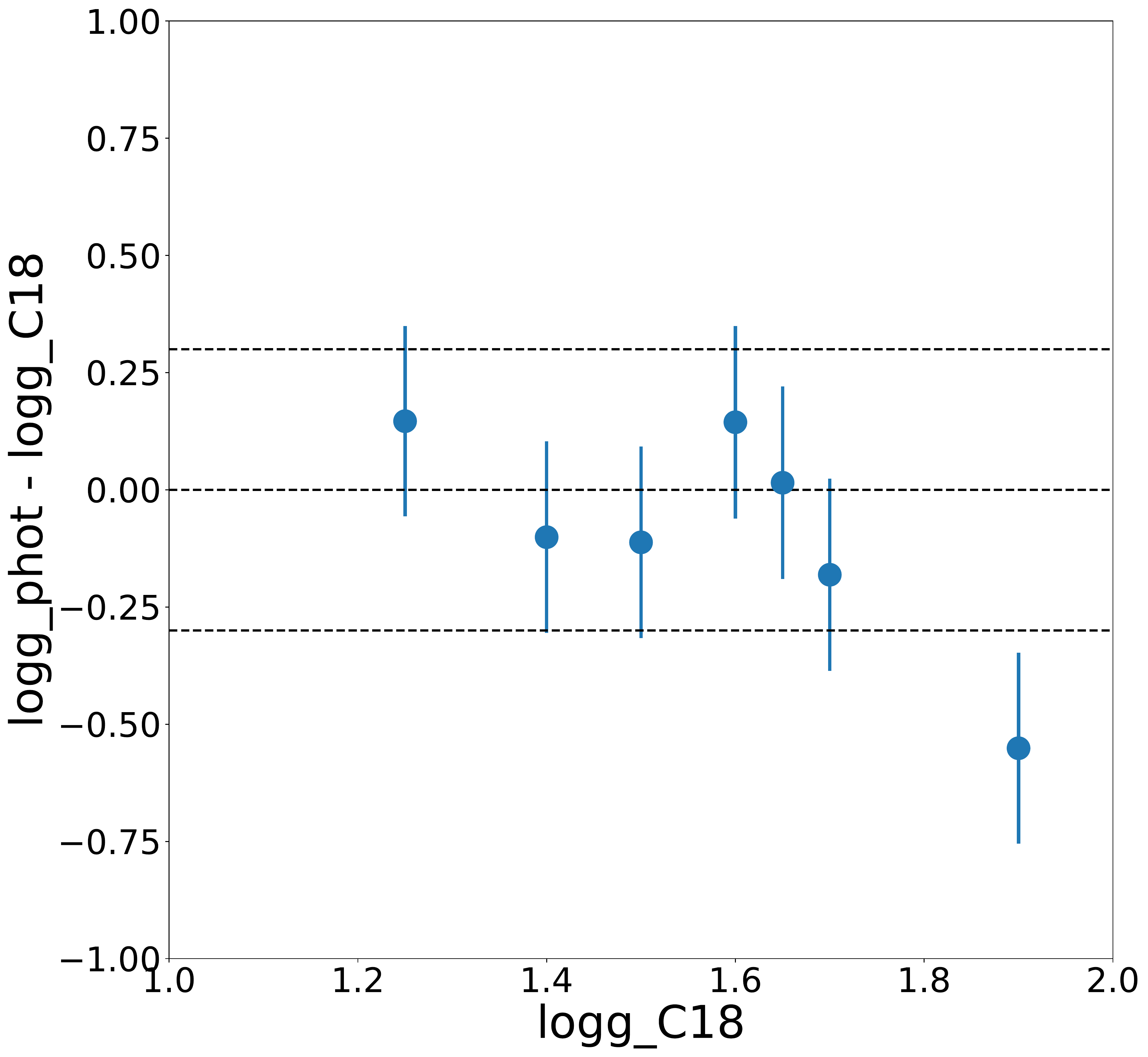}
\caption{Comparison of our photometric $\log g$ values to those derived from high-resolution spectroscopy in \citet{cfj+18}.
Dashed lines are drawn at $+0.3$ and $-0.3$ to guide the eye.
Error bars correspond to the uncertainty in the photometric $\log g$ values of these stars.}
\label{fig:logg_compare}
\end{figure}

\subsection{Comparison to Tuc II high-resolution members}
\label{sec:hires}

Seven stars in Tucana II have high-resolution spectroscopic metallicities presented in \citet{cfj+18}.
We compare our photometric metallicities and surface gravities for those seven stars to the spectrosopically determined values.
The results are shown in the left panel of Figure~\ref{fig:feh_compare} and in Figure~\ref{fig:logg_compare}. 

For the metallicities, we find a mean offset between our values of 0.09\,dex (in which we measure a higher [Fe/H]) with a standard deviation of 0.34\,dex.
This is excellent agreement, given that all but one photometric metallicity is within $1\sigma$ agreement of the metallicities derived from high-resolution spectroscopy (see Section~\ref{sec:sigmas} for a discussion of the derivation of these uncertainties).
Additionally, the mean carbon abundance of these stars is $\mbox{[C/Fe]}=0.35$ and their mean $T_{\text{eff}}$ is 4870\,K, which, according to Figure~\ref{fig:Carbon} would suggest that we should overestimate the photometric metallicity by $\sim$0.1\,dex. 

We also find generally excellent agreement between our photometric $\log g$ values and those in \citet{cfj+18}, as shown in Figure~\ref{fig:logg_compare}.
The mean offset is $-$0.09\,dex, meaning we measure a lower $\log g$ relative to those derived from high-resolution spectroscopy.
The standard deviation of the residual between our $\log g$ measurements is 0.22\,dex. 
The standard deviation and mean offset are almost entirely driven by the one outlier, labeled as TucII-078 in \citet{cfj+18}, at $\log g \sim 1.9$ in Figure~\ref{fig:logg_compare}. 
Excluding it would change the offset to +0.02\,dex and the standard deviation to 0.11\,dex.
However, the presence of this outlier is not entirely surprising, given that \citet{cfj+18} derive a correspondingly large uncertainty of 0.67\,dex in the $\log\,g$ of TucII-078.

\subsection{Comparison to \citet{wmo+16}}
\label{sec:medres}

We also compare our photometric metallicities to [Fe/H] values from \citet{wmo+16}, who derived these values from $R\sim18,000$ and $R\sim10,000$ spectra of the Mg b region ($\sim5150$\AA) for 137 candidate member stars in the vicinity of Tucana\,II.
We chose to only compare with stars in \citet{wmo+16} that had metallicity uncertainties $< 0.20$\,dex to ensure a high-quality comparison.
We further only compared to stars with [Fe/H] $< -1.0$ in \citet{wmo+16}, as our grid of synthetic photometry only extends to [Fe/H] $= -0.5$.
We find good agreement between our photometric metallicity measurements and those in \citet{wmo+16}, with a mean offset of $-$0.12\,dex, meaning we measure a lower [Fe/H], and a standard deviation between our measurements of 0.23\,dex.

\subsection{Final [Fe/H] and $\log{g}$ uncertainties}
\label{sec:sigmas}

We assume that the uncertainty in each photometric [Fe/H] value is a combination of 1) intrinsic uncertainty from our methodology and 2) random uncertainty that is propagated from uncertainties in the photometry.
The intrinsic uncertainty in our methodology is assumed to be 0.16\,dex, which is the standard deviation of the residuals of our photometric metallicities for globular clusters (see Section~\ref{sec:GCs}).
We thus take into account that the mean photometric [Fe/H] value for each cluster is usually derived from a large number of stars ($N > 10$), suggesting that the standard error in each of these values is generally small.
We therefore assume that the $0.16$\,dex scatter in the residuals is mostly driven by the intrinsic uncertainty in our method, which we adopt as such when calculating the final uncertainty in our photometric [Fe/H] values.
The random uncertainty is derived by adding in quadrature the difference in photometric [Fe/H] obtained after varying each the $v$, $g$, and $i$ magnitudes by their $1\sigma$ photometric uncertainties and redetermining final values.
If the variation of any of the magnitudes by their $1\sigma$ photometric uncertainties takes them beyond the bounds of the grid of synthetic photometry, then a conservative uncertainty of 0.75\,dex is adopted.

The intrinsic and random uncertainties are then added in quadrature to derive final uncertainties on our photometric [Fe/H] values.
Our final photometric metallicity uncertainties appear to be reasonable, as the median of the uncertainty of the photometric metallicities in the left and right panels of Figure~\ref{fig:feh_compare} is 0.31\,dex and 0.21\,dex, respectively.
These uncertainties are similar to the standard deviations of the data points in each of these panels of 0.34\,dex and 0.23\,dex, respectively.
For another estimate of the precision of our method, we can thus pool together the residuals in Figures~\ref{fig:GCmetallicities} and~\ref{fig:feh_compare} and compute their standard deviation.
Upon doing this, we find that the standard deviation of the residuals is 0.20\,dex for data points with [Fe/H] $> -2.5$ and 0.34\,dex for those with [Fe/H] $< -2.5$.

The uncertainty in the photometric $\log g$ is calculated a similar manner.
The random uncertainty is derived by adding in quadrature the difference in photometric $\log g$ after varying the $u$, $v$, $g$, and $i$ magnitudes by their $1\sigma$ photometric uncertainties.
The intrinsic uncertainty is assumed to be 0.20\,dex, since this leads to the median uncertainty of the photometric $\log g$ values in Figure~\ref{fig:logg_compare} to agree with the standard deviation of the residuals.

\section{Re-discovering the Tucana II dwarf galaxy}
\label{sec:TucII}

In this section, we show that our photometric metallcities and surface gravities, when combined with \textit{Gaia} DR2 proper motion data, provide an extremely efficient means to identify likely member stars (Section~\ref{sec:i_members}) of Tucana II. This principally enables precise studies of the properties of a UFD.
We further outline a method using the python \texttt{emcee} package to quantify membership probability (Section~\ref{sec:p_members}) to derive a metallicity distribution function (MDF) for the Tucana II UFD (Section~\ref{sec:fehdist}). 
In principle, these techniques could also be used to  study other UFDs with incomplete or no spectroscopy of member stars.

\begin{figure*}[!htbp]
\centering
\includegraphics[width =\textwidth]{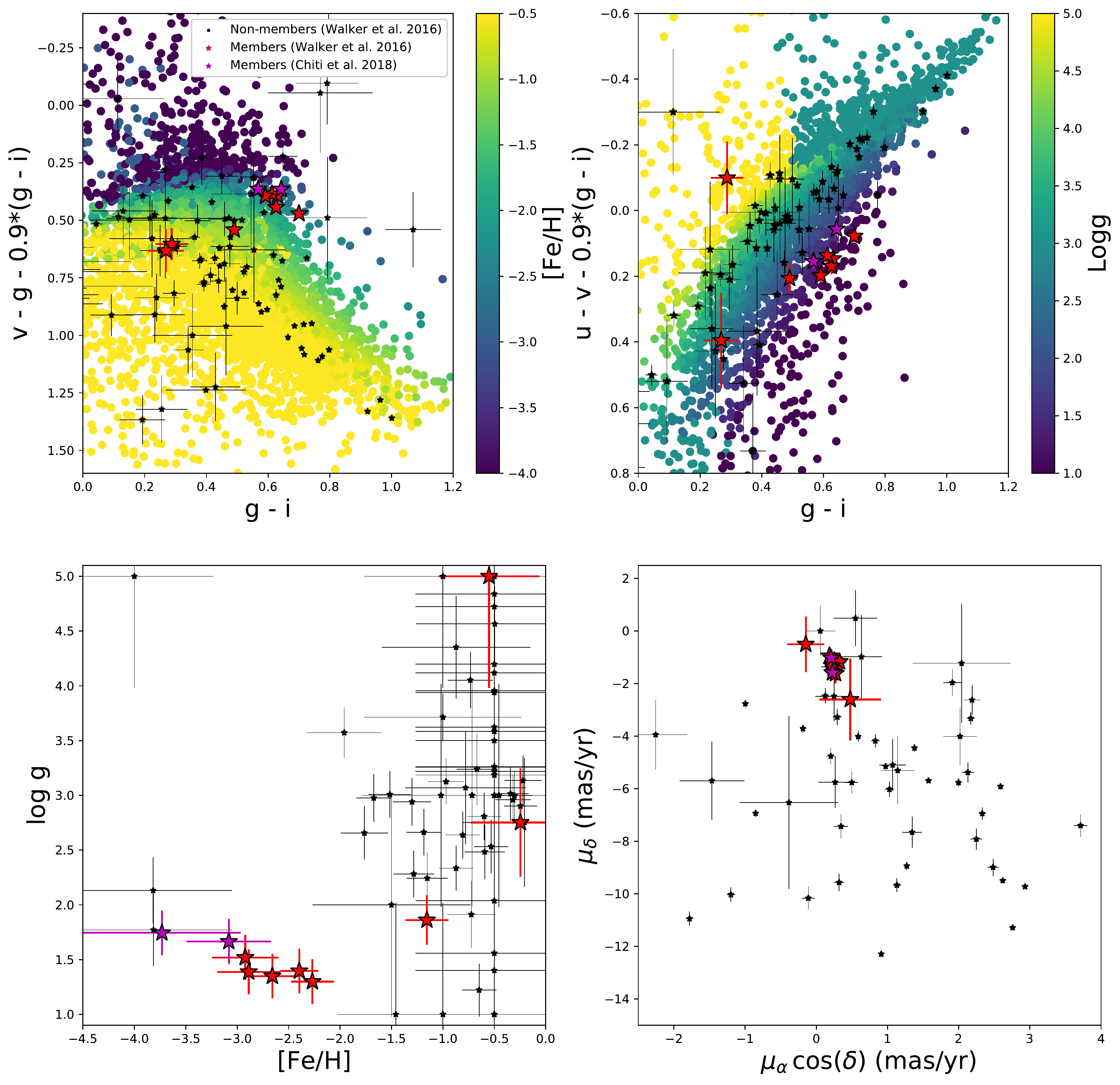}
\caption{Top left: Color-color plot for determining photometric metallicities.
All stars  selected along an isochrone that matches the Tucana II red giant branch stellar population are colored according to their photometric metallcities.
Red star symbols are confirmed members (p $>$ 0.95) from \citet{wmo+16}; small black star symbols are confirmed non-members from \citet{wmo+16}; purple star symbols are confirmed members from high-resolution spectroscopy from \citet{cfj+18}.
Top right: Color-color plot for determining photometric $\log g$.
As discussed in Section~\ref{sec:p_members}, stars with [Fe/H] $> -1.0$ and $\log g \geq 3.0$ are excluded when deriving membership probabilities since they are very likely to be foreground contaminants.
Bottom left: A plot of photometric $\log g$ vs. photometric [Fe/H] for stars in \citet{wmo+16} and \citet{cfj+18}. 
Note the separation of the majority of confirmed members from foreground stars at high metallicites.
Bottom right: \textit{Gaia} DR2 proper motions of stars in \citet{wmo+16} and \citet{cfj+18}. }
\label{fig:find_members}
\end{figure*}

\subsection{Identifying members of Tucana II}
\label{sec:i_members}

We performed several preliminary steps to prepare our source catalog for analysis.
We first removed galaxies by cross-matching our sources with those in the DES Y1A1 gold catalog \citep{dsr+18}.
Following the criteria and procedure described in \citet{dam+12} and \citet{bdb+15}, we excluded all sources from the DES catalog with the parameter \texttt{SPREAD\_MODEL\_I} $>$ 0.003 \citep{dam+12} to retain only stars.
We then measured photometric metallicities and surface gravities of every star within the parameters of our grid of synthetic photometry, and compiled their proper motion measurements from the \textit{Gaia} DR2 catalog \citep{gaia+16, gaia+18}. Then, we compiled a list of confirmed member stars of Tucana II from \citet{wmo+16} and \citet{cfj+18} and confirmed non-member stars in the vicinity of Tucana II from \citet{wmo+16}.
\citet{wmo+16} derived membership probabilities for 137 stars in the vicinity of the Tucana II UFD. 
Two new confirmed member stars of Tucana II had already been identified by \citet{cfj+18} from the data presented in this paper.

We consider all stars with a membership probability $>$ 95\% in \citet{wmo+16} to be likely members in our subsequent analysis. Of particular interest in this regard is whether likely member stars could be separated from non-member stars using our photometric stellar parameter measurements and \textit{Gaia} proper motion data.
In Figure~\ref{fig:find_members}, we illustrate several tests to qualitatively separate the non-members and likely members from \citet{wmo+16} and \citet{cfj+18}.
The top two panels demonstrate that the majority of likely members separate from non-members in the color-color plots we employ to measure photometric stellar parameters. 
The bottom two panels of Figure~\ref{fig:find_members} show that when combining metallicity, $\log g$, and proper motion information, the confirmed members are largely distinct from foreground stars.
However, three members from \citet{wmo+16} do not separate as cleanly.
One of these stars is fairly metal-rich ([Fe/H] $\sim-1.3$), about 1 dex more compared to the other member stars. 
It is likely a metal-rich, or carbon-enhanced (see Section~\ref{sec:carbon}), member of the UFD, given its separation from the foreground in $\log g$ and proper motion.
The other two stars may indeed be non-members, given our measurements of their photometric metallicity ([Fe/H] $> -1)$ and surface gravities ($\log\,g > 2.5$).
Furthermore, the slight separation of these two stars from the other confirmed member stars in proper motion space, as shown in the bottom right panel of Figure~\ref{fig:find_members}, supports this notion.
However, since these two stars are faint ($g >20$) and their measurements are correspondingly less precise, firm arguments about their membership status cannot be made.

In conclusion, however, we demonstrate that photometric metallicities and surface gravities, especially in the case of high-quality measurements, can clearly separate UFD member stars from foreground stars.
Furthermore, the \textit{Gaia} proper motion data is useful in identifying likely members that may otherwise be metal-rich (i.e., the one star at [Fe/H] $\sim -1.3$ in the bottom left plot of Figure~\ref{fig:find_members}).

\begin{figure*}[!htbp]
\centering
\includegraphics[width =\textwidth]{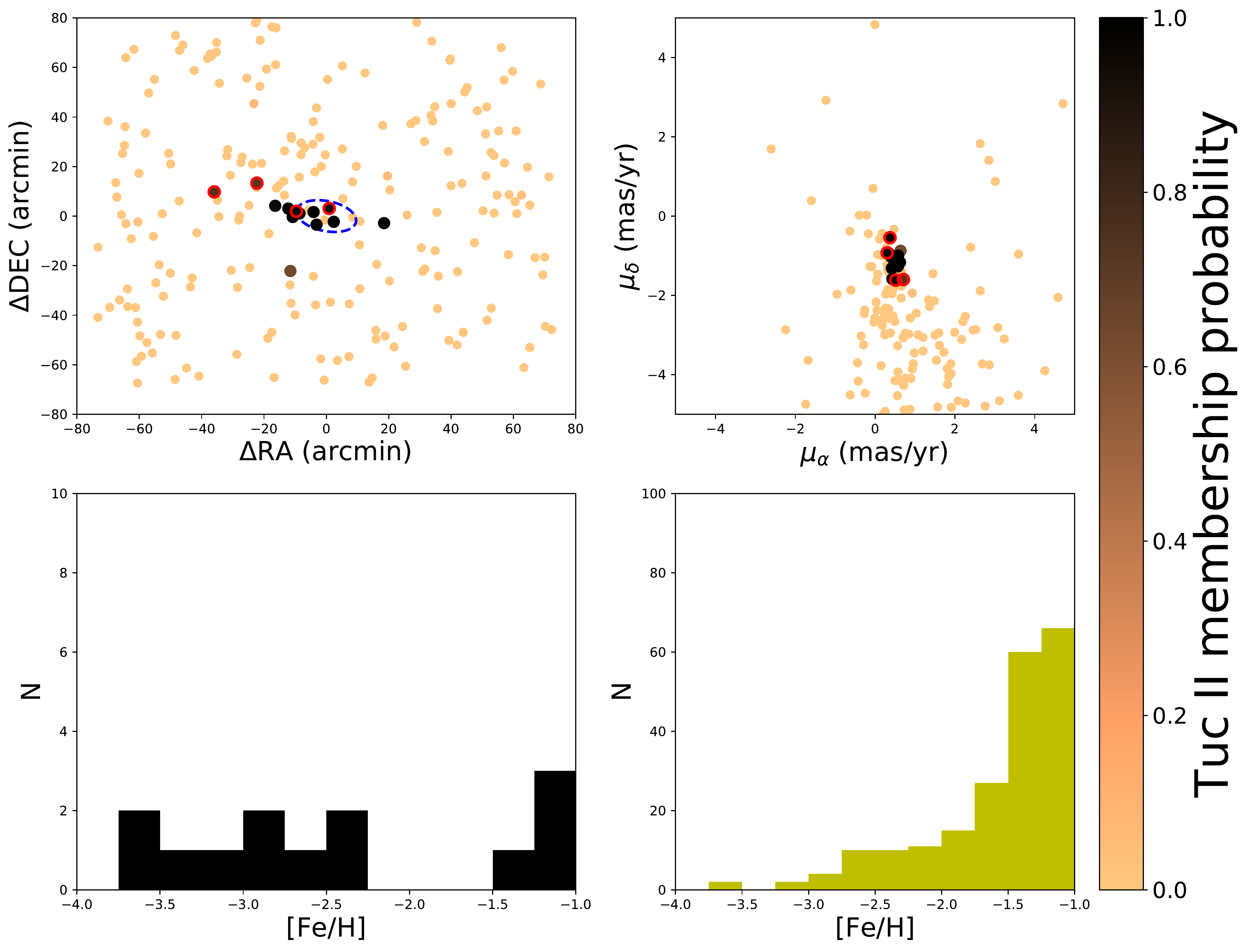}
\caption{Top left: Location of each metal-poor ([Fe/H] $< -1.0$) giant ($\log g <$ 3.0) in our sample of stars, and colored by membership probability.
As expected, we identify a number of likely members near the center of the galaxy. The three likely members with photometric [Fe/H] $> -$1.5 are circled in red. The half-light radius from \citet{kbt+15} is overplotted in blue. 
Top right: Same as the top left, but plotted in proper-motion space. 
We find that the likely Tucana II members are tightly clustered in proper-motion space which is unsurprising given the small intrinsic dispersion in the system. 
Bottom panels: Metallicities of stars with membership probability $p > 0.50$ (left) and $p < 0.50$ (right). 
Despite not applying an additional metallicity-dependent term in calculating the membership probabilities, beyond the initial sample cut we find that the metallicities of the likely members are on average more metal-poor than the likely non-members. See text for discussion.}
\label{fig:p_mems}
\end{figure*}

\subsection{Quantifying membership probabilities}
\label{sec:p_members}

In order to quantitatively derive properties (i.e., MDF) of the Tucana II UFD, member stars need to be selected well despite the presence of large numbers of foreground stars.
Given that we derive photometric $\log\,g$ and [Fe/H] values, we can immediately remove foreground metal-rich, and main sequence stars from our sample to alleviate the issue of significant foreground contamination.
Then, as previously demonstrated in e.g., \citet{pl+19}, we use a combination of the spatial location of each star and its \textit{Gaia} DR2 proper motion measurements to derive quantitative membership probabilities for each star.
The bottom panels of Figure~\ref{fig:find_members} already qualitatively show that using photometric metallicity, $\log g$, and proper motions enable adequate membership identification.

To quantify the membership likelihood of each star, we then
proceed with several steps.
First, we remove stars that are either metal-rich ([Fe/H] $> -1.0$), not on the red giant branch ($\log g \geq 3.0$), or fainter than $g=20$. 
We apply the brightness cut to ensure that our stars have reliable photometric $\log g$ and [Fe/H] values. 
We apply the metallicity cut since no metal-rich stars are known as members of UFDs (see i.e., \citealt{fn+15}, \citealt{s+19} for reviews), and we apply the $\log g$ cut as stars in Tucana II down to $g=20$ are at the base of the red giant branch or higher up.

We model the remaining set of metal-poor giants using a mixture model with the following likelihood function:
\begin{eqnarray}
\label{eqn:total}
\Lagr_{\text{Total}} = f_{\text{mem}}\Lagr_{\text{sp,mem}}\Lagr_{\text{pm,mem}} + 
\nonumber\\ (1 - f_{\text{mem}})\Lagr_{\text{sp,nonmem}}\Lagr_{\text{pm,nonmem}}
\end{eqnarray}
where $f_{\text{mem}}$ denotes the fraction of member stars of Tucana II. 
The spatial distribution of member stars of Tucana II is assumed to follow an exponential profile, following \citet{mjh+08} and \citet{lms+19}, with a likelihood function given by:
\begin{eqnarray}
\Lagr_{\text{sp,mem}} = \left.\frac{\exp(-\frac{R}{R_e})}{2\pi R_e(1-\epsilon)} \right/ \int_S \frac{\exp(-\frac{R}{R_e})}{2\pi R_e(1-\epsilon)}\,dS
\end{eqnarray}
where $\epsilon$ is the ellipticity, $R_e$ is the exponential radius, and R is the elliptical radius, given by:
\begin{eqnarray}
R=((\frac{1}{1-\epsilon}((x-x_0)\cos\theta - (y-y_0)\sin\theta))^2\nonumber\\
+ ((x-x_0)\sin\theta +(y-y_0)\cos\theta)^2)^{1/2}
\end{eqnarray}
where $x_0$ and $y_0$ are the right ascension and declination of the center of Tucana II, as measured in \citet{kbt+15}, and $\theta$ is the position angle of the elliptical distribution.
$x$ and $y$ are the distances from the center of Tucana II along the direction of right ascension and declination, respectively.
The spatial distribution of non-members, $\Lagr_{\text{sp,nonmem}}$ is assumed to be uniform.

The likelihood functions for the proper motions of members and non-members, $\Lagr_{\text{pm,mem}}$ and  $\Lagr_{\text{pm,nonmem}}$ are assumed to be bivariate Gaussians, following the formalism presented in e.g., \citet{lms+19}.
We use a standard bivariate gaussian to model the foreground stars, but use the following probability density to model the members of Tucana II:
\begin{eqnarray}
\Scale[1.025]{p = \frac{(2\pi)^{-1}}{\sigma_{\mu_\alpha}\,\sigma_{\mu_\delta}} \times} \nonumber\\
\Scale[1.025]{\exp \left[-\frac{(\mu_{\alpha} - \left<\mu_{\alpha}\right>_{\text{TII}})^2} {2\sigma^2_{\mu_\alpha}}-\frac{(\mu_{\delta}-\left<\mu_{\delta}\right>_{\text{TII}}-k\,(\mu_{\alpha}-\left<\mu_{\alpha}\right>_{\text{TII}}))^2}{2\sigma^2_{\mu_\delta}}\right]}
\end{eqnarray}
where $\mu_{\alpha}$ is \textit{Gaia} proper motion in the direction of right ascension, $\mu_{\delta}$ is the proper motion is the direction of declination, $\sigma_{\mu_\alpha}$ and $\sigma_{\mu_\delta}$ are their corresponding uncertainties, $\left<\mu_{\alpha}\right>_{\text{TII}}$ and $\left<\mu_{\alpha}\right>_{\text{TII}}$ denote the systemic proper motion of the Tucana II UFD, and $k$ adds the analog of a position angle to the bivariate gaussian.
Identically to \citet{pl+19}, we assume that the intrinsic proper motion uncertainties are much smaller than the observational uncertainties in proper motion, and thus the width of the bivariate Gaussian for the UFD members is solely determined by the uncertainties in proper motions.

We sample the likelihood function using the \texttt{emcee} package \citep{fhl+13}, which uses the ensemble sampler from \citet{gw+10}.
There were 8 free parameters in our sampling, which were $f_\text{mem}$, $k$, $\left<\mu_{\alpha}\right>_{\text{TII}}$, $\left<\mu_{\alpha}\right>_{\text{TII}}$, $\left<\mu_{\alpha}\right>_{\text{MW}}$, $\left<\mu_{\alpha}\right>_{\text{MW}}$, $\sigma_{\mu_{\alpha,\text{MW}}}$, $\sigma_{\mu_{\delta,\text{MW}}}$.
The parameters $R_e$, $\epsilon$, $\theta$ were fixed to the values provided in \citet{kbt+15}.
We initialized the sampler with 200 walkers, with 2000 steps after a burn-in period of 500 steps to ensure a good sampling of the posterior distributions. 
The membership probability was then simply assumed to be the ratio of the membership terms to the total likelihood in equation~\ref{eqn:total}.

\begin{figure*}[!htbp]
\centering
\includegraphics[width =\textwidth]{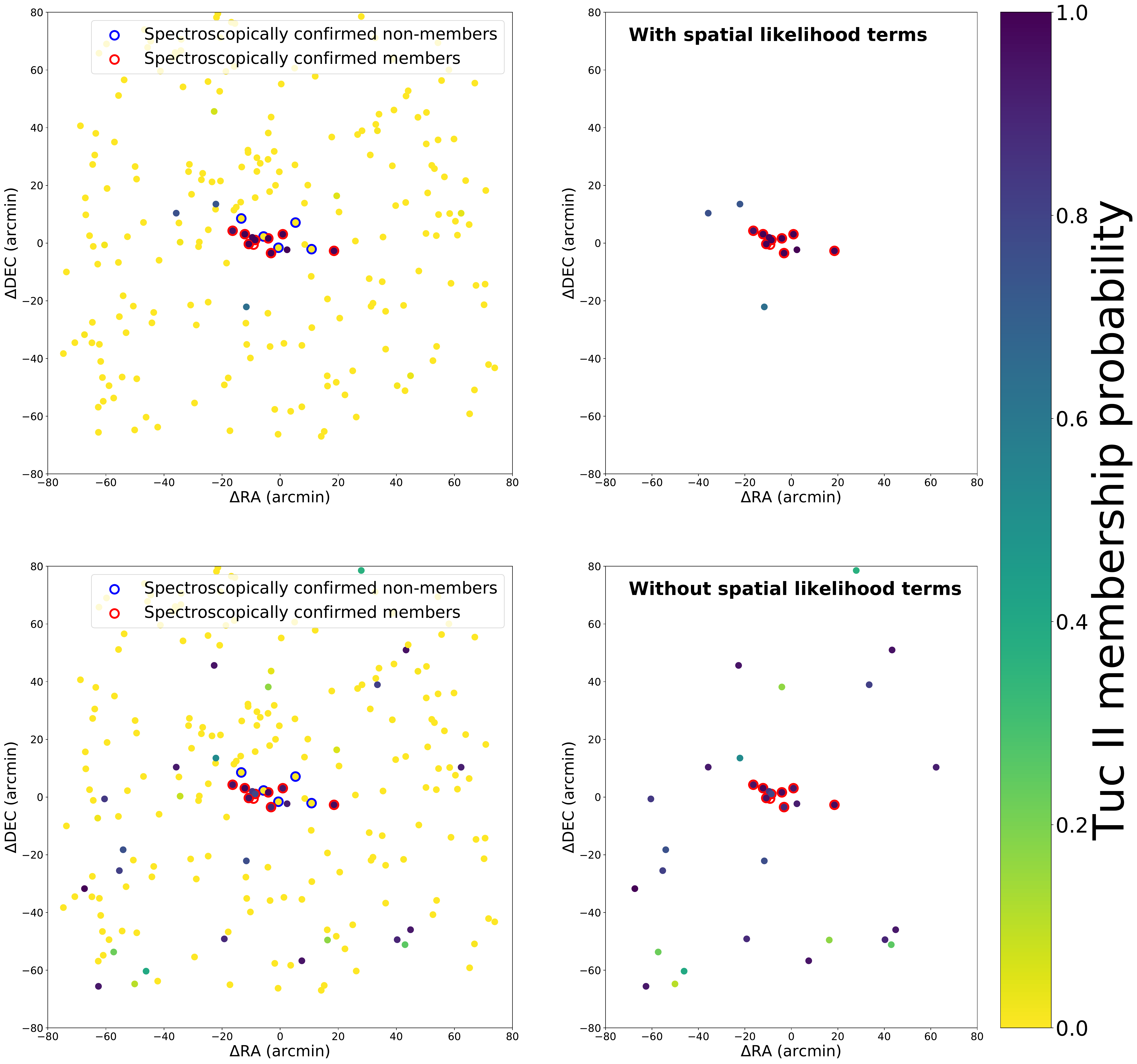}
\caption{Top left: Spatial distribution of stars, colored by their membership probability using the likelihood function in Equation~\ref{eqn:total}. 
Stars with membership probabilities $p > 0.95$ and $g < 20$ in \citet{wmo+16}, as well as additional confirmed members from \citet{cfj+18}, are outlined in red.
Stars with membership probabilities $p < 0.50$ and $g < 20$ in \citet{wmo+16} are outlined in blue.
Top right: Same as left panel, but only including stars with membership probability of $p > 0.10$ from our study.
We find that we exclude all known non-members and recover all known members in the literature, except for one horizontal branch star, which naturally would have been excluded by our selection along the red giant branch of an isochrone.
Bottom panels: Same as top panels, but membership probabilities are computed excluding the spatial terms in the likelihood function in Equation~\ref{eqn:total}.
We find several additional candidate member stars in Tucana II that are distant from the center of the system.
However, further investigation is needed before these distant stars can be classified as likely members.
Note that the color scheme for membership probabilities is different compared to that used in Figure~\ref{fig:p_mems}, to visually aid the identification of marginal members.}
\label{fig:p_mems_spatial}
\end{figure*}

In Table~\ref{tab:TucII}, Figure~\ref{fig:p_mems}, and the top panels of Figure~\ref{fig:p_mems_spatial}, we present our final membership probability for each star.
As expected, we find a number of likely members near the nominal center of Tucana II. 
However, interestingly, we also find several stars well separated from the center of the galaxy that also have a likelihood of being members. 

To ensure that our identification of likely members is accurate, we investigated whether we recovered the known sample of Tucana II members from \citet{wmo+16} and \citet{cfj+18}.
We only considered stars with DES $g < 20$ in \citet{wmo+16}, as this is comparable to the initial magnitude cut for our sample.
We consider stars that have membership probability $p > 0.95$ in \citet{wmo+16} and the two additional members presented in \citet{cfj+18} to be likely members, and stars with membership probability $p < 0.50$ in \citet{wmo+16} to be likely non-members.
Results of our comparison are shown in Figure~\ref{fig:p_mems_spatial}.

As an additional check, we also compared our catalog of likely members with that of \citet{pl+19}, who identified likely members based on \textit{Gaia} DR2 proper motions and DES photometry.
We find that we recover their entire sample of 10 likely ($p > 0.50$) members of the red giant branch of Tucana II brighter than $g\sim19.6$.
We additionally identify three likely member stars not found as having membership probability greater than 0.50 in their catalog. 
We note that two of our likely members that appear to be more metal-rich ([Fe/H] $> -1.5$) also appear as likely members in \citet{pl+19}, further supporting that they are indeed members.

We find that our initial exclusion of stars with photometric [Fe/H] $> -1.0$ and $\log g > 3.0$ removes all but five likely non-members from \citet{wmo+16}.
Furthermore, the five remaining likely non-members are identified as likely non-members from our method, since we derive membership probabilities $p < 0.10$ for all those stars.
We re-identify all likely members of the red giant branch from \citet{wmo+16} and \citet{cfj+18} as highly likely members in our sample (all $p > 0.99$).
The one likely member we are not recovering from \citet{wmo+16} is on the horizontal giant branch, since we exclude stars not on the red giant branch from our sample.
In addition to re-identifying all the known likely members on the red giant branch, we here identify likely members both in the core of Tucana II and several half-light radii away from it. 
This result demonstrates that SkyMapper photometry and \textit{Gaia} proper motions very efficiently identify likely member stars of UFDs, and by extension, alleviate the problem of foreground contamination when studying these systems.

We further note that our selection procedure excludes all known non-members in the core of Tucana\,II.
This fact implies that it may be possible to identify members of Tucana II, agnostic of the spatial distribution of stars.
Purely as an exercise, we recompute membership probabilities after excluding the spatial terms in Equation~\ref{eqn:total} and present the result in the bottom panels of Figure~\ref{fig:p_mems_spatial}.
We find, as expected, that we still exclude all known non-members in the core of Tucana II and re-identify the likely members on the red giant branch. 
We additionally find a number of candidate members that are many half-light radii from the center of the system.
In upcoming work, we indeed confirm the membership status of a handful of these distant stars (A. Chiti et al., in prep), which is suggestive of a more spatially extended population of stars, some tidal disturbance, or a need to revisit the structural parameters of the system.

\begin{figure*}[!htbp]
\centering
\includegraphics[width =\textwidth]{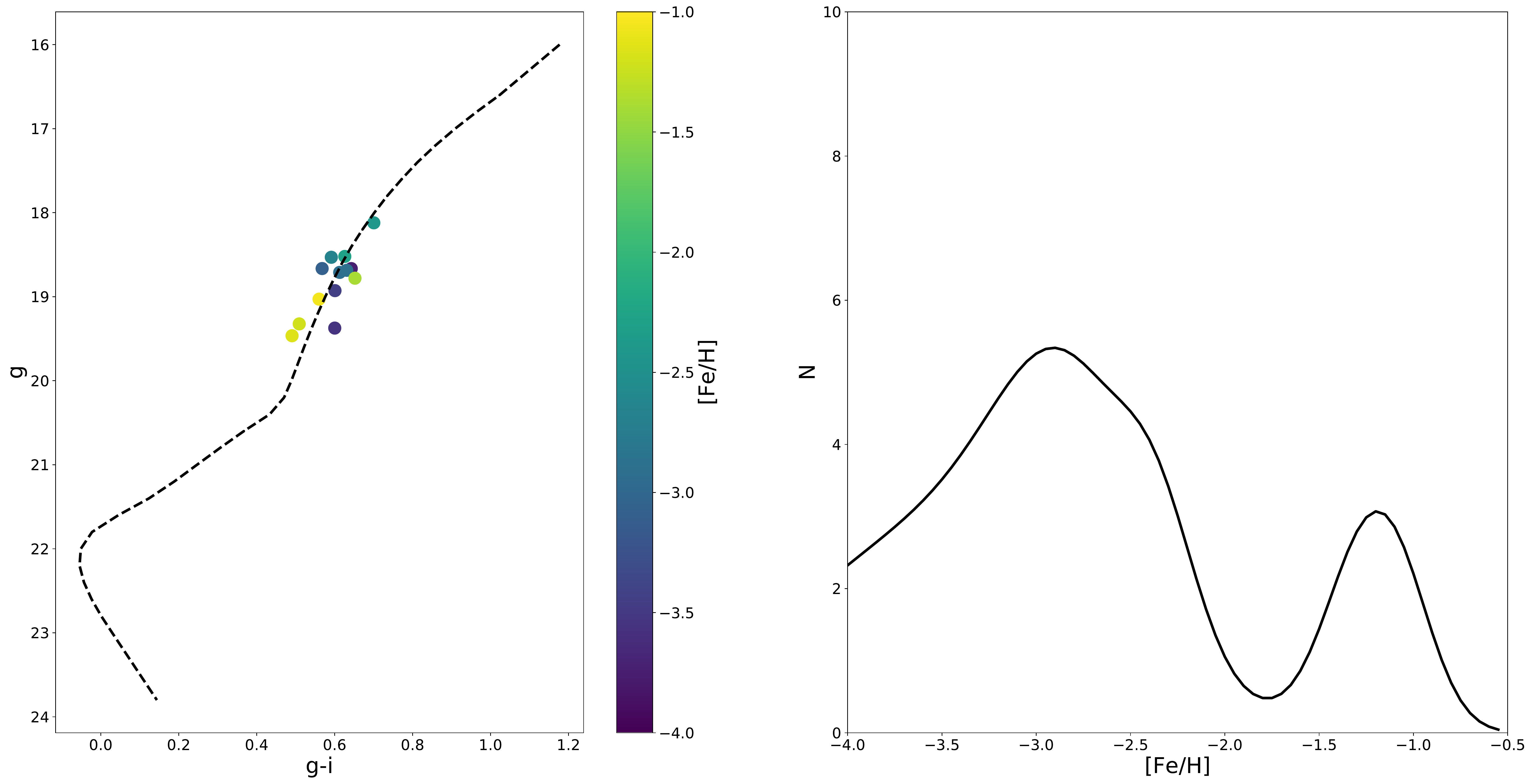}
\caption{Left: color-magnitude diagram of Tucana II with stars that have membership probability $p > 0.50$, based on the methodology presented in this paper.
Each star is colored by its photometric metallicity value.
A 12 Gyr, [Fe/H] = $-2.5$ MIST isochrone is overplotted for reference \citep{d+16, cdc+16}.
Right: metallicity distribution function (MDF) of the Tucana II UFD based on our membership likelihood analysis.
It is largely composed of extremely metal-poor stars with $\mbox{[Fe/H]}\sim-3$ but we also find a population of stars at photometric [Fe/H]$\sim-1.25$. The latter group is suggestive of additional carbon-enhanced metal-poor giants with overestimated metallicities (see Section~\ref{sec:carbon} for discussion).
We note that all stars with photometric [Fe/H] $> -1.0$ are removed from our sample, and our grid for deriving metallicities extends down to [Fe/H] = $-4.0$.
Thus, our MDF is only populated by stars with metallicities between those values, and extensions above [Fe/H] = $-1.0$ and below [Fe/H] = $-4.0$ are due to uncertainties in the photometric [Fe/H] for individual stars.}
\label{fig:MDF}
\end{figure*}

\subsection{Metallicity distribution of likely Tucana II members}
\label{sec:fehdist}

Given the membership probabilities obtained using Equation~\ref{eqn:total} in Section~\ref{sec:p_members}, we can now derive a MDF for  Tucana II including spatial priors.
We compile the photometric metallicity values and uncertainties derived in Section~\ref{sec:analysis}, together with our membership probabilities (see Section~\ref{sec:p_members}). 
We then generate a Gaussian for each star in which the mean is the photometric metallicity, with one $\sigma$ being equal to the uncertainty in the photometric metallicity, and the amplitude being equal to the membership probability. 
We then simply sum these Gaussians to generate a MDF. 
The result is shown in Figure~\ref{fig:MDF}. 

We find a population of extremely metal-poor stars in this distribution which makes Tucana II one of the most metal-poor galaxies, with a MDF peaking at [Fe/H]$\sim-2.9$. This result follows earlier investigations that also yielded overall low metallicities for the system \citep{jfe+16, cfj+18}.
However, we also find a more metal-rich component around [Fe/H]$\sim-1.25$.
If these photometric [Fe/H] values are taken at face value, this higher metallicity component would suggest an extended formation history for the system. 
Any indications for an extended star formation history might, however, suggest that even more stars at higher metallicity are present in the system.
This has not been found by other studies, as all other UFDs  have been shown to not contain stars with $\mbox{[Fe/H]}>-1.0$ \citep{s+19}. 
It is thus unlikely that the apparent bump at [Fe/H]$\sim-1.25$ represents a truly metal-rich component.
As discussed in detail in Section~\ref{sec:carbon}, the absorption features around the Ca II K line (most importantly the CN feature) artificially increase the measured metallicity of each star in accordance with their carbon abundance. Thus, the high metallicity population may instead be indicative of the presence of strongly carbon-enhanced CEMP stars in Tucana II. Given that the metal-poor halo population contains a significant fraction of CEMP stars \citep{pfb+14} and that UFDs have not yielded many strongly enhanced CEMP stars ([C/Fe] $> 1.0$), this is an interesting option to explore further with spectroscopic followup observations. 


\begin{deluxetable*}{ccccccc}
\tablecolumns{7}
\tablewidth{\columnwidth}
\tablecaption{\label{tab:TucII} Photometric metallicities of stars with Tucana II membership probabiliy $p_{\text{mem}} > 0.50$}
\tablehead{
\colhead{RA (deg) (J2000)} & 
\colhead{DEC (deg) (J2000)} & 
\colhead{$g_{\text{SM}}$} & 
\colhead{$i_{\text{SM}}$} & 
\colhead{$p_{\text{mem}}$} & 
\colhead{[Fe/H]$_{\text{phot}}$} & 
\colhead{$\sigma$([Fe/H]$_{\text{phot}}$)}}
\startdata
342.671075 & $-$58.518976 & 18.58 & 17.94 & 1.00 & $-$2.66 & 0.26 \\
342.959507 & $-$58.627823 & 18.17 & 17.42 & 1.00 & $-$2.40 & 0.19 \\
343.136343 & $-$58.608469 & 19.42 & 18.77 & 1.00 & $-$3.56 & 0.75 \\
342.929412 & $-$58.542702 & 18.73 & 18.06 & 1.00 & $-$2.89 & 0.31 \\
343.089087 & $-$58.518710 & 19.51 & 18.97 & 1.00 & $-$1.16 & 0.21 \\
342.753839 & $-$58.537255 & 19.37 & 18.81 & 0.99 & $-$1.21 & 0.20 \\
342.784619 & $-$58.552258 & 18.57 & 17.90 & 0.99 & $-$2.27 & 0.21 \\
342.715112 & $-$58.575714 & 18.76 & 18.10 & 0.99 & $-$2.92 & 0.32 \\
343.652659 & $-$58.616101 & 18.71 & 18.10 & 0.99 & $-$3.08 & 0.41 \\
342.537105 & $-$58.499739 & 18.72 & 18.02 & 0.96 & $-$3.73 & 0.75 \\
341.916360 & $-$58.402040 & 19.09 & 18.47 & 0.75 & $-$1.05 & 0.19 \\
342.352876 & $-$58.346508 & 18.83 & 18.13 & 0.73 & $-$1.39 & 0.18 \\
342.687905 & $-$58.939023 & 18.98 & 18.33 & 0.64 & $-$3.44 & 0.75
\enddata
\end{deluxetable*}


\section{Conclusion}
\label{sec:conclusion}

In this paper, we present an application of deep imaging carried out with the SkyMapper telescope to derive stellar parameters for stars.
We find that by modeling the predicted fluxes for stars over a wide range of stellar parameters, we can accurately and precisely measure the $\log g$ and metallicities of stars, solely from  photometry.

We then apply this technique to the Tucana II UFD.
Previous studies of UFDs were hampered by the presence of foreground, metal-rich main-sequence stars. 
We demonstrate that by leveraging these photometric stellar parameters, we can efficiently identify member stars of these generally metal-poor systems and derive photometric metallicities for these stars. 
We can also derive quantitative membership probabilities by using \textit{Gaia} DR2 proper motion data, after removing foreground contaminants using our photometric stellar parameters.

Using these membership probabilities, we are able to (1) identify a handful of stars several half-light radii from the center of Tucana II that have high membership probabilities and (2) derive a MDF for the system.
We identify additional possible members of the Tucana II UFD upon removing the spatial likelihood terms when computing membership probabilities.
Follow-up spectroscopy of several of these stars will be presented in an upcoming paper (A. Chiti et al, in prep), in which we find that a handful of distant stars are indeed members, based on spectroscopic metallicities and radial velocity measurements.
The MDF of Tucana II is either suggestive of an extended period of star formation history, or the presence of some very carbon-enhanced metal-poor stars in Tucana II, with the latter option being more likely.
Future work will apply this technique to other UFDs to derive spatially complete MDFs and to investigate whether any other UFDs may host a spatially extended population of stars.

\acknowledgements
We thank Christian Wolf, Chris Onken, and Dougal Mackey for helpful discussions on the SkyMapper data reduction process. 
A.C. and A.F. are partially supported by NSF- CAREER
grant AST-1255160 and NSF grant 1716251. 
H.J. acknowledges support from the Australian Research Council through the Discovery Project DP150100862.
This work made
use of NASAs Astrophysics Data System Bibliographic
Services, and the SIMBAD database, operated at CDS,
Strasbourg, France \citep{woe+00}.
This work has made use of the VALD database, operated at Uppsala University, the Institute of Astronomy RAS in Moscow, and the University of Vienna. 

This work has made use of data from the European Space Agency (ESA) mission
{\it Gaia} (\url{https://www.cosmos.esa.int/gaia}), processed by the {\it Gaia}
Data Processing and Analysis Consortium (DPAC,
\url{https://www.cosmos.esa.int/web/gaia/dpac/consortium}). Funding for the DPAC
has been provided by national institutions, in particular the institutions
participating in the {\it Gaia} Multilateral Agreement.

The national facility capability for SkyMapper has been funded through ARC LIEF grant LE130100104 from the Australian Research Council, awarded to the University of Sydney, the Australian National University, Swinburne University of Technology, the University of Queensland, the University of Western Australia, the University of Melbourne, Curtin University of Technology, Monash University and the Australian Astronomical Observatory. SkyMapper is owned and operated by The Australian National University's Research School of Astronomy and Astrophysics.
This research has made use of the SVO Filter Profile Service (http://svo2.cab.inta-csic.es/theory/fps/) supported from the Spanish MINECO through grant AYA2017-84089

Facilities: SkyMapper \citep{ksb+07}

\software{Turbospectrum \citep{ar+98, p+12}, MARCS \citep{gee+08}, Astropy \citep{astropy}, NumPy \citep{numpy}, SciPy \citet{jop+01}, Matplotlib \citep{Hunter+07}}

\bibliography{tucii_phot}

\end{document}